\documentclass[journal,subeqn,caption]{IEEEtran}
\usepackage{mathtools}
\usepackage{algorithm}
\usepackage{algpseudocode}
\usepackage{hhline}
\usepackage{pdflscape}
\usepackage{balance}

\usepackage{cite}
\usepackage{float}
\usepackage{multicol}
\usepackage{multirow}
\usepackage{balance}
\usepackage{graphicx}
\usepackage{amssymb}
\usepackage{cuted}

\usepackage{rotating,booktabs,dcolumn}
\usepackage[flushleft]{threeparttable}

\DeclarePairedDelimiter\abs{\lvert}{\rvert}%
\DeclarePairedDelimiter\norm{\lVert}{\rVert}%

\makeatletter
\let\oldabs\abs
\def\abs{\@ifstar{\oldabs}{\oldabs*}}
\let\oldnorm\norm
\def\norm{\@ifstar{\oldnorm}{\oldnorm*}}
\makeatother

\usepackage{array}

\newcolumntype{P}[1]{>{\centering\arraybackslash}p{#1}}
\newcolumntype{M}[1]{>{\centering\arraybackslash}m{#1}}

\usepackage[hidelinks]{hyperref}
\usepackage{graphicx}
\usepackage{amsfonts}
\usepackage{amsmath}
\usepackage{mathabx}
\usepackage{bm}
\usepackage{color}
\usepackage{epstopdf}
\usepackage{mathtools, nccmath}
\usepackage{multirow}
\usepackage{bbm}
\usepackage{marginnote}
\usepackage{tikz}
\usetikzlibrary{shapes,arrows,chains}
\usepackage{enumitem}
\usepackage{comment}
\usepackage{eurosym}
 \usepackage{epstopdf}
\usetikzlibrary{arrows,decorations.pathmorphing,backgrounds,fit,positioning,shapes.symbols,chains}
\usepackage{caption}
\usepackage{colortbl}
\usepackage{float}
\usepackage{xcolor}
\usepackage[caption=false]{subfig}
\usepackage{tabularx}

\definecolor{Gray}{gray}{0.9}
\definecolor{LightCyan}{rgb}{0.88,1,1}

\newcolumntype{a}{>{\columncolor{Gray}}c}
\newcolumntype{b}{>{\columncolor{white}}c}

\makeatletter
\newcases{Rcases}{$\mskip\thickmuskip$}{%
  \hfil$\m@th{##}$}{$\m@th{##}$\hfil}{.}{\rbrace}
\makeatother
\allowdisplaybreaks

\usepackage{diagbox} 
\usepackage{longtable}
\usepackage{array}
\usepackage{rotating}
\usepackage{eqparbox}
\usepackage{makecell, caption, booktabs}
\DeclarePairedDelimiter\ceil{\lceil}{\rceil}

 \ifCLASSINFOpdf
\else
\fi

\begin{document}
\newcommand{\AF}{E}
\newcommand{\AT}{E^\mathrm{R}}
\newcommand{\A}{\AF \!\cup \!\AT}
\newcommand{\BP}{N^{\mathrm{P}}}
\newcommand{\BL}{L_i}
\newcommand{\BS}{S_i}
\newcommand{\G}{G}
\newcommand{\BG}{G_i}
\newcommand{\RB}{R}

\newcommand{\BPR}{N^{\mathrm{PR}}}

\newcommand{\BB}{\beta}

\newcommand{\Xiul}{\Xi^{\mathrm{ul}}}
\newcommand{\Xill}{\Xi^{\mathrm{ll}}}
\newcommand{\Xir}{\Xi^{\mathrm{r}}}
\newcommand{\Xidu}{\Xi^{\mathrm{du}}}

\renewcommand{\wp}{OP}
\newcommand{\es}{es}

\newcommand{\CC}{\bm{\ddot{c}}_{k}}
\newcommand{\C}{\bm{\dot{c}}_{k}}
\newcommand{\cc}{\bm{c}_{k}}
\newcommand{\Pgen}{P^{\mathrm{g}}_{t,k}}
\newcommand{\Qgen}{Q^{\mathrm{g}}_{t,k}}
\newcommand{\Pgenm}{P^{\mathrm{g}}_{t-1,k}}
\newcommand{\Qgenm}{Q^{\mathrm{g}}_{t-1,k}}
\renewcommand{\P}{P_{t,e,i,j}}
\newcommand{\Q}{Q_{t,e,i,j}}
\newcommand{\Sline}{\bm{\overline{S}}_{e}}
\newcommand{\dP}{\bm{P}^{\mathbf{d}}_{t,l}}
\newcommand{\dQ}{\bm{Q}^{\mathbf{d}}_{t,l}}
\newcommand{\Vtn}{\bm{V}^{\mathbf{\wp}}_{t,i}}
\newcommand{\Vtm}{\bm{V}^{\mathbf{\wp}}_{t,j}}
\newcommand{\fitn}{\bm{\theta}^{\mathbf{\wp}}_{t,i}}
\newcommand{\fitm}{\bm{\theta}^{\mathbf{\wp}}_{t,j}}
\newcommand{\dVn}{V^{\Delta}_{t,i}}
\newcommand{\dVm}{V^{\Delta}_{t,j}}
\newcommand{\dfin}{\theta^{\Delta}_{t,i}}
\newcommand{\dfim}{\theta^{\Delta}_{t,j}}

\newcommand{\Va}{\widecheck{V}_{t,e}}

\newcommand{\cosf}{\widehat{cos}_{t,i,j}}
\newcommand{\cost}{\widehat{cos}_{t,j,i}}

\newcommand{\gl}{\bm{g}_{e}}
\newcommand{\gf}{\bm{g}^{\mathbf{fr}}_{e}}
\newcommand{\gt}{\bm{g}^{\mathbf{to}}_{e}}

\newcommand{\bl}{\bm{b}_{e}}
\renewcommand{\bf}{\bm{b}^{\mathbf{fr}}_{e}}
\newcommand{\bt}{\bm{b}^{\mathbf{to}}_{e}}

\newcommand{\gs}{\bm{g}^{\mathbf{sh}}_{s}}
\newcommand{\bs}{\bm{b}^{\mathbf{sh}}_{s}}

\newcommand{\tap}{\bm{\tau}_{e}}
\newcommand{\shift}{\bm{\sigma}_{e}}

\newcommand{\Pmax}{\overline{\bm{P}}^{\mathbf{g}}_k}
\newcommand{\Pmin}{\underline{\bm{P}}^{\mathbf{g}}_k}
\newcommand{\Qmax}{\overline{\bm{Q}}^{\mathbf{g}}_k}
\newcommand{\Qmin}{\underline{\bm{Q}}^{\mathbf{g}}_k}
\newcommand{\Vmax}{\overline{\bm{V}}_{i}}
\newcommand{\Vmin}{\underline{\bm{V}}_{i}}

\newcommand{\cospsinij}{\bm{cps}_{t,e,i,j}}
\newcommand{\cospsinji}{\bm{cps}_{t,e,j,i}}
\newcommand{\cosmsinij}{\bm{cms}_{t,e,i,j}}
\newcommand{\cosmsinji}{\bm{cms}_{t,e,j,i}}

\newcommand{\vpJedan}{\bm{p}_{1,t,e,i,j}}
\newcommand{\vpDva}{\bm{p}_{2,t,e}}
\newcommand{\vpTri}{\bm{p}_{3,t,e,i,j}}

\newcommand{\condv}{\bm{\Lambda}_{t,e}}
\newcommand{\condc}{\bm{\Gamma}_{t,i,j}}
\newcommand{\condcji}{\bm{\Gamma}_{t,j,i}}
\newcommand{\conds}{\bm{\Phi}_{t,e,i,j}}

\newcommand{\SoEmax}{\overline{\bm{SoE}}}
\newcommand{\etach}{\bm{\eta}^{\mathbf{ch}}}
\newcommand{\etadis}{\bm{\eta}^{\mathbf{dis}}}
\newcommand{\qchmax}{\bm{\overline{s}}^{\mathbf{\es}}}
\newcommand{\qdismax}{\bm{\overline{s}}^{\mathbf{\es}}}

\newcommand{\SoE}{SoE_{t}}
\newcommand{\SoEm}{SoE_{t-1}}
\newcommand{\qbat}{p^{\mathrm{\es}}_{t}}
\newcommand{\qch}{p^{\mathrm{ch}}_{t}}
\newcommand{\qdis}{p^{\mathrm{dis}}_{t}}
\newcommand{\qqbat}{q^{\mathrm{\es}}_{t}}
\newcommand{\qqch}{q^{\mathrm{ch}}_{t}}
\newcommand{\qqdis}{q^{\mathrm{dis}}_{t}}

\newcommand{\xch}{x^{\mathrm{p}}_{t}}
\newcommand{\xxch}{x^{\mathrm{q}}_{t}}

\newcommand{\lambdaJedan}{\lambda_{1,t,i}}
\newcommand{\lambdaJedanDn}{\underline{\bm{\lambda}}_{1,t,i}}
\newcommand{\lambdaJedanUp}{\overline{\bm{\lambda}}_{1,t,i}}

\newcommand{\lambdaDva}{\lambda_{2,t,i}}
\newcommand{\lambdaDvaDn}{\underline{\bm{\lambda}}_{2,t,i}}
\newcommand{\lambdaDvaUp}{\overline{\bm{\lambda}}_{2,t,i}}

\newcommand{\lambdaTri}{\lambda_{3,t,e,i,j}}
\newcommand{\lambdaTriji}{\lambda_{3,t,e,j,i}}
\newcommand{\lambdaCetiri}{\lambda_{4,t,e,i,j}}
\newcommand{\lambdaCetiriji}{\lambda_{4,t,e,j,i}}
\newcommand{\lambdaPet}{\lambda_{5,t,e,i,j}}
\newcommand{\lambdaPetji}{\lambda_{5,t,e,j,i}}
\newcommand{\lambdaSest}{\lambda_{6,t,e,i,j}}
\newcommand{\lambdaSestji}{\lambda_{6,t,e,j,i}}
\newcommand{\lambdaSedam}{\lambda_{16,t,i}}

\newcommand{\lambdaJedanaest}{\lambda_{14,t,e,i,j}}
\newcommand{\lambdaDvanaest}{\lambda_{15,t,e,i,j}}
\newcommand{\lambdaTrinaest}{\mu_{5,t,e,i,j}}

\newcommand{\lambdaCetrnaest}{\lambda_{11,t,i,j}}
\newcommand{\lambdaPetnaest}{\lambda_{12,t,i,j}}
\newcommand{\lambdaSesnaest}{\mu_{2,t,i,j}}
\newcommand{\lambdaCetrnaestji}{\lambda_{11,t,j,i}}

\newcommand{\lambdaSedamnaest}{\lambda_{7,t,e,i,j}}
\newcommand{\lambdaOsamnaest}{\lambda_{8,t,e,i,j}}
\newcommand{\lambdaOsamnaestji}{\lambda_{8,t,e,j,i}}
\newcommand{\lambdaDevetnaest}{\lambda_{9,t,e,i,j}}
\newcommand{\lambdaDvadeset}{\mu_{1,t,e,i,j}}

\newcommand{\lambdaDvadesetjedan}{\lambda_{10,t,e,i,j}}
\newcommand{\lambdaDvadesetdva}{\lambda_{13,t,i,j}}

\newcommand{\muJedanDn}{\underline{\mu}_{3,t,k}}
\newcommand{\muJedanUp}{\overline{\mu}_{3,t,k}}
\newcommand{\muDvaDn}{\underline{\mu}_{4,t,k}}
\newcommand{\muDvaUp}{\overline{\mu}_{4,t,k}}
\newcommand{\muTriDn}{\underline{\mu}_{6,t,i}}
\newcommand{\muTriUp}{\overline{\mu}_{6,t,i}}

\newcommand{\cvJedan}{f_{1,t,i,j}}
\newcommand{\cvDva}{f_{2,t,i,j}}
\newcommand{\cvTri}{f_{0,t,i,j}}

\newcommand{\vvJedan}{w_{1,t,e,i,j}}
\newcommand{\vvDva}{w_{2,t,e,i,j}}
\newcommand{\vvTri}{w_{3,t,e,i,j}}
\newcommand{\vvCetiri}{w_{0,t,e,i,j}}

\newcommand{\Omegad}{\Omega^{\mathrm{d}}}
\newcommand{\Omegap}{\Omega^{\mathrm{p}}}

\newcommand{\xnula}{x_{0}}
\newcommand{\xjedan}{x_{1}}
\newcommand{\xdva}{x_{2}}
\newcommand{\xtri}{x_{3}}
\newcommand{\xcrta}{\overline{x}}
\newcommand{\x}{x}

\newcommand{\ynula}{y_{0}}
\newcommand{\yjedan}{y_{1}}
\newcommand{\ydva}{y_{2}}
\newcommand{\ytri}{y_{3}}
\newcommand{\ycrta}{\overline{y}}
\newcommand{\y}{y}

\newcommand{\eps}{\bm{\epsilon}}
\newcommand{\smf}{F}
\newcommand{\psin}{\psi_{n}}
\newcommand{\psijedan}{\psi_{1}}
\newcommand{\psidva}{\psi_{2}}
\newcommand{\un}{u_{n}}
\newcommand{\ujedan}{u_{1}}
\newcommand{\udva}{u_{2}}

\newcommand{\wbat}{w^{\mathrm{p}}_{t}}
\newcommand{\wch}{w^{\mathrm{p,ch}}_{t}}
\newcommand{\wdis}{w^{\mathrm{p,dis}}_{t}}

\newcommand{\wwbat}{w^{\mathrm{q}}_{t}}
\newcommand{\wwch}{w^{\mathrm{q,ch}}_{t}}
\newcommand{\wwdis}{w^{\mathrm{q,dis}}_{t}}

\newcommand{\ybat}{y^{\mathrm{p}}_{t}}
\newcommand{\yybat}{y^{\mathrm{q}}_{t}}

\newcommand{\ib}{b}
\newcommand{\iu}{u}
\newcommand{\xbin}{x^{\mathrm{p,be}}_{t,\ib}}
\newcommand{\xxbin}{x^{\mathrm{q,be}}_{t,\ib}}

\newcommand{\US}{U}
\newcommand{\BSS}{B}
\newcommand{\wbe}{w^{\mathrm{p,be}}_{t,\ib}}
\newcommand{\wwbe}{w^{\mathrm{q,be}}_{t,\ib}}
\newcommand{\wue}{w^{\mathrm{p,ue}}_{t,\iu}}
\newcommand{\wwue}{w^{\mathrm{q,ue}}_{t,\iu}}
\newcommand{\xun}{x^{\mathrm{p,ue}}_{t,\iu}}
\newcommand{\xxun}{x^{\mathrm{q,ue}}_{t,\iu}}

\newcommand{\pen}{\bm{\pi}}



\title{Solving Bilevel AC OPF Problems by Smoothing the Complementary Conditions -- Part II: Solution Techniques and Case Study}
\author{K. \v{S}epetanc, \textit{Student Member}, \textit{IEEE}, H. Pand\v{z}i\'c, \textit{Senior Member}, \textit{IEEE} and T. Capuder, \textit{Member}, \textit{IEEE}
%
%
\thanks{The authors are with the Innovation Centre Nikola Tesla (ICENT) and the University of Zagreb Faculty of Electrical Engineering and Computing (e-mails: karlo.sepetanc@fer.hr; hrvoje.pandzic@fer.hr; tomislav.capuder@fer.hr). Employment of Karlo \v{S}epetanc is funded by the Croatian Science Foundation under programme DOK-2018-09. The research leading to these results has received funding from the European Union’s Horizon 2020 research and innovation programme under grant agreement No 864298 (project ATTEST). The sole responsibility for the content of this document lies with the authors. It does not necessarily reflect the opinion of the Innovation and Networks Executive Agency (INEA) or the European Commission (EC). INEA or the EC are not responsible for any use that may be made of the information contained therein.}
}
%
%
\maketitle

\begin{abstract}
This is a second part of the research on AC optimal power flow being used in the lower level of the bilevel strategic bidding or investment models. As an example of a suitable upper-level problem, we observe a strategic bidding of energy storage and propose a novel formulation based on the smoothing technique.

After presenting the idea and scope of our work, as well as the model itself and the solution algorithm in the companion paper (Part I), this paper presents a number of existing solution techniques and the proposed one based on smoothing the complementary conditions. The superiority of the proposed algorithm and smoothing techniques is demonstrated in terms of accuracy and computational tractability over multiple transmission networks of different sizes and different OPF models. The results indicate that the proposed approach outperforms all other options in both metrics by a significant margin. This is especially noticeable in the metric of accuracy where out of total 422 optimizations over 9 meshed networks the greatest AC OPF error is 0.023\% that is further reduced to 3.3e-4\% in the second iteration of our algorithm.


%
\end{abstract}

\section{Introduction}
\label{sec:intro}

In the Part I of this work, we develop and present a mathematical formulation of a bilevel problem based on the smoothing techniques, where a strategic player’s profit maximization is in the upper level and the AC OPF approximation in the lower level.
Building upon Algorithm I presented in the Part I paper, in this paper we present a number of solution techniques that can be used to solve the bilevel strategic bidding problem at hand, as well as any other bilevel problem with the AC optimal power flow (AC OPF)-based market clearing algorithm in the lower level. These solution techniques, described in Section \ref{sec:sol_tech} of this paper, act as a baseline to which we compare the proposed smoothing techniques which differ in the smoothing function. The case study Section \ref{sec:case} consists of three main parts: the description and set-up Subsection \ref{sub:description}; and the two case studies. The first case study described in \ref{sub:accuracy} demonstrates the model's accuracy, while the second one in \ref{sub:tractability} presents an in-depth solution techniques analysis. The final Section IV provides conclusive remarks.

\section{Solution Techniques}\label{sec:sol_tech}

In this work we consider all classical techniques to solve a single-level reduced bilevel optimization problem, i.e. Step 5 of the Algorithm 1 from the Part I paper. The techniques differ in how well they close the duality gap, numerical tractability and ease of finding or converging to the global optimality as opposed to a local one. In the following subsections we first present the classical techniques, i.e. the primal-dual counterpart, the strong duality, the McCormic envelopes, the complementarity slackness, the penalty factor, the interaction discretization, followed by the proposed smoothing techniques, Chen-Harker-Kanzow-Smale and Kanzow.



\subsection{Primal-dual counterpart}\label{sub:PD}

\subsubsection{Primal-dual (PD)}\hfill

This technique relies on the convexified objective function \eqref{eq:ofqp} to act as a penalty factor and does not enforce closure of the duality gap in any other way. One of the first such convexifications is presented in \cite{PD}. The resulting model simply consists of the lower-level primal and dual constraints as well as the upper-level constraints and the convexified quadratic objective function. Using this technique, the problem belongs to the second-order cone programming (SOCP) optimization class (with convex quadratic objective function). Dual constraints are available in the Appendix of the Part I paper.


The upper-level objective function is convexified by adding the term $\Omegad - \Omegap$. This way, the bilinear terms that cause nonconvexity, $\qbat \cdot \lambdaJedan$ and $\qqbat \cdot \lambdaDva$, are canceled since they also appear in the dual objective function $\Omegad$. The convexified objective function is equivalent to the original one if zero duality gap is ensured by the solution technique, i.e. $\Omegad=\Omegap$.

\begin{equation}\tag{4.1}
\mathrm{Max}\:\: \sum_{t,i\in \BB} (\qbat \!\cdot\! \lambdaJedan \!+\!\qqbat \!\cdot\! \lambdaDva) + \Omegad - \Omegap
\label{eq:ofqp}
\end{equation}

The problem maximizes \eqref{eq:ofqp} subject to constraints (1.2)--(1.7), (2.2)--(2.14), excluding (2.8.1) and (2.9.1) in favor of (3.1.1)--(3.2.4) and (B.2)--(B.13), with respect to the variables set $\Xiul \cup \Xir \cup \Xidu$. Note that equations (1)--(3) and (B) are from the Part I paper, where (1) denote the upper-level constraints, constraints (2) lower-level primal constraints (primal feasibility KKT conditions), (3) are reformulated lower-level SOC constraints, i.e. reformulated constraints (2), while (B) are lower-level dual constraints, i.e. stationarity and dual feasibility KKT conditions.

\subsubsection{Strengthened primal-dual (PD-S)}\hfill

This technique applies an additional linear constraint \eqref{eq:PD.1} on top of the regular primal-dual technique. This constraint is obtained by writing the Karush–Kuhn–Tucker (KKT) stationarity conditions for $\Pgen$. Otherwise, it is used to derive the dual as a part of dealing with the objective function convex quadraticity by substituting $\Pgen$ from the Langrange function. Therefore, this constraint is not directly a part of the dual model, but it closes the duality gap since it connects both the primal and the dual variables. This technique is applicable only if there are generators with square cost bids ($\CC>0$). The optimization class is SOCP.

\begin{equation}\tag{4.2}
\C + \muJedanUp - \muJedanDn + \!\!\! \sum_{\substack{i \\ :k \in \BG}}\!\! \lambdaJedan + 2 \cdot \CC \cdot \Pgen = 0, \quad \forall{t,k}: \CC>0
\label{eq:PD.1}
\end{equation}

The problem maximizes \eqref{eq:ofqp} subject to constraints (1.2)--(1.7), (2.2)--(2.14), excluding (2.8.1) and (2.9.1) in favor of (3.1.1)--(3.2.4), (B.2)--(B.13) and \eqref{eq:PD.1}, with respect to the variables set $\Xiul \cup \Xir \cup \Xidu$.

\subsection{Strong Duality}\label{sub:SD}
\subsubsection{Strong Duality (SD)}\hfill

Strong duality technique directly enforces zero duality gap by enforcing constraint \eqref{eq:SD.1}, as explained in \cite{SD}. The formulation is nonconvex quadratic due to the equality sign and bilinear $\qbat \cdot \lambdaJedan$ and $\qqbat \cdot \lambdaDva$ terms in eq. \eqref{eq:SD.1}.

\begin{equation}\tag{4.3}
\Omegap=\Omegad
\label{eq:SD.1}
\end{equation}

The problem maximizes \eqref{eq:ofqp} subject to constraints (1.2)--(1.7), (2.2)--(2.14), excluding (2.8.1) and (2.9.1) in favor of (3.1.1)--(3.2.4), (B.2)--(B.13) and \eqref{eq:SD.1}, with respect to the variables set $\Xiul \cup \Xir \cup \Xidu$.

\subsubsection{Relaxed Strong Duality (SD-R)}\hfill

To potentially improve numerical stability, the strong duality constraint can be relaxed so that small gaps are allowed. For any primal-dual optimization problem pair that has a finite solution, assuming that the goal of the primal is minimization, $\Omegap\ge\Omegad$ weak duality holds even without such constraint in the model. Thus, it is sufficient to add a constraint with an opposite inequality sign to close the duality gap. Constraint \eqref{eq:SD.2} allows an absolute duality gap $\eps$, where $\eps$ is a small positive constant. The formulation is nonconvex quadratic due to bilinear $\qbat \cdot \lambdaJedan$ and $\qqbat \cdot \lambdaDva$ terms in constraint \eqref{eq:SD.2}.



\begin{equation}\tag{4.4}
\Omegap \le \Omegad +\eps
\label{eq:SD.2}
\end{equation}

The problem maximizes \eqref{eq:ofqp} subject to constraints (1.2)--(1.7), (2.2)--(2.14), excluding (2.8.1) and (2.9.1) in favor of (3.1.1)--(3.2.4), (B.2)--(B.13) and \eqref{eq:SD.2}, with respect to the variables set $\Xiul \cup \Xir \cup \Xidu$.

\subsection{McCormick Envelopes (MC)}\label{sub:MC}


McCormick envelopes \cite{MC} relax a bilinear term into a plane-bounded region. The technique requires an assumption on the bounds of the electricity price, i.e. $\lambdaJedanDn$, $\lambdaDvaDn$ for the lower bound and $\lambdaJedanUp$, $\lambdaDvaUp$ for the upper bound. Values of these parameters can be estimated from the obtained operating point in Step 1 of Algorithm 1, e.g. using fixed intervals around the computed prices. Variables $\wbat$ and $\wwbat$ represent relaxations of $\qbat \cdot \lambdaJedan$ and $\qqbat \cdot \lambdaDva$ respectively. Constraint \eqref{eq:MC.1} is the strong duality constraint in the inequality form. $+\qbat \cdot \lambdaJedan$ and $+\qqbat \cdot \lambdaDva$ terms from \eqref{eq:MC.1} cancel the original negative terms from $\Omegad$, which are replaced with $-\wbat$ and $-\wwbat$. Constraints \eqref{eq:MC.2} and \eqref{eq:MC.3} are McCormick underestimator planes, while constraints \eqref{eq:MC.4} and \eqref{eq:MC.5} are McCormick overestimator planes. Together, they form a relaxed convex feasible space. To shorten the formulation, they are written in the matrix form. The formulation belongs to the SOCP optimization class.

\begin{equation}\tag{4.5}
\Omegap \le \Omegad +  \hspace{-2pt} \sum_{t,i\in \BB} \hspace{-2pt} \qbat \!\cdot \! \lambdaJedan +  \hspace{-2pt} \sum_{t,i\in \BB} \hspace{-2pt} \qqbat \!\cdot \! \lambdaDva - \sum_{t}\wbat - \sum_{t}\wwbat
\label{eq:MC.1}
\end{equation}

\begin{equation}\tag{4.6}
\begin{split}
\begin{pmatrix}\wbat &\!\!\!\! \wwbat \end{pmatrix}^{\intercal} \!\!\ge \!\! &-\!\qdismax \!\cdot\!\begin{pmatrix}\lambdaJedan & \!\!\!\! \lambdaDva\end{pmatrix}^{\intercal} \!+\! \begin{pmatrix}\qbat& \!\!\!\! \qqbat \end{pmatrix}^{\intercal} \!\cdot\! \begin{pmatrix}\lambdaJedanDn & \!\!\!\!\lambdaDvaDn \end{pmatrix} \\
&+\! \qdismax \!\cdot\! \begin{pmatrix} \lambdaJedanDn & \!\!\!\! \lambdaDvaDn \end{pmatrix}^{\intercal}, \quad \forall{t,i\in \BB}\\[-0.95cm]
\label{eq:MC.2}
\end{split}
\end{equation}

\begin{equation}\tag{4.7}
\begin{split}
\begin{pmatrix}\wbat &\!\!\!\! \wwbat \end{pmatrix}^{\intercal} \!\ge  &\qchmax \!\cdot\! \begin{pmatrix}\lambdaJedan & \!\!\!\! \lambdaDva\end{pmatrix}^{\intercal} \!+\! \begin{pmatrix}\qbat& \!\!\!\! \qqbat \end{pmatrix}^{\intercal} \!\cdot\! \begin{pmatrix}\lambdaJedanUp & \!\!\!\!\lambdaDvaUp \end{pmatrix} \\
- &\qchmax \!\cdot\! \begin{pmatrix} \lambdaJedanUp & \!\!\!\! \lambdaDvaUp \end{pmatrix}^{\intercal}, \quad \forall{t,i\in \BB}\\[-0.95cm]
\label{eq:MC.3}
\end{split}
\end{equation}

\begin{equation}\tag{4.8}
\begin{split}
\begin{pmatrix}\wbat &\!\!\!\! \wwbat \end{pmatrix}^{\intercal}  \!\le & \qchmax \!\cdot\! \begin{pmatrix}\lambdaJedan & \!\!\!\! \lambdaDva\end{pmatrix}^{\intercal} \!+\! \begin{pmatrix}\qbat& \!\!\!\! \qqbat \end{pmatrix}^{\intercal} \!\cdot\! \begin{pmatrix}\lambdaJedanDn & \!\!\!\!\lambdaDvaDn \end{pmatrix} \\
- &\qchmax \!\cdot\! \begin{pmatrix} \lambdaJedanDn & \!\!\!\! \lambdaDvaDn \end{pmatrix}^{\intercal}, \quad \forall{t,i\in \BB}\\[-0.95cm]
\label{eq:MC.4}
\end{split}
\end{equation}

\begin{equation}\tag{4.9}
\begin{split}
\begin{pmatrix}\wbat &\!\!\!\! \wwbat \end{pmatrix}^{\intercal} \!\le \!\! & -\qdismax \!\cdot\! \begin{pmatrix}\lambdaJedan & \!\!\!\! \lambdaDva\end{pmatrix}^{\intercal} \!+\! \begin{pmatrix}\qbat& \!\!\!\! \qqbat \end{pmatrix}^{\intercal} \!\cdot\! \begin{pmatrix}\lambdaJedanUp & \!\!\!\!\lambdaDvaUp \end{pmatrix} \\
&+ \qdismax \!\cdot\! \begin{pmatrix} \lambdaJedanUp & \!\!\!\! \lambdaDvaUp \end{pmatrix}^{\intercal}, \quad \forall{t,i\in \BB}\\[-0.95cm]
\label{eq:MC.5}
\end{split}
\end{equation}

The problem maximizes \eqref{eq:ofqp} subject to constraints (1.2)--(1.7), (2.2)--(2.14), excluding (2.8.1) and (2.9.1) in favor of (3.1.1)--(3.2.4), (B.2)--(B.13) and \eqref{eq:MC.1}--\eqref{eq:MC.5}, with respect to the variables set $\Xiul \cup \Xir \cup \Xidu \cup \{\wbat,\wwbat\}$.

\subsection{Complementary Slackness}\label{sub:CC}

\subsubsection{Complementary Slackness (CS)}\hfill

For any given basic SOC primal-dual constraint inequality pair \eqref{eq:CC.1} and \eqref{eq:CC.2}, assuming primal vector variable $x \!=\!\begin{pmatrix}\xnula & \xcrta\end{pmatrix}$, where $\xcrta\!=\!\begin{pmatrix}\xjedan & \xdva & \xtri &...\end{pmatrix}$, and analogous dual vector variable $y$, there is a complementary slackness condition \eqref{eq:CC.3}. In case of linear inequalities, $\xcrta$ is an empty vector. Thus, for linear inequalities, constraints \eqref{eq:CC.1}--\eqref{eq:CC.3} take the following forms respectively: $0 \! \le \! \xnula$, $0 \! \le \! \ynula$ and $\xnula \!\cdot\! \ynula \!=\! 0$. Normally, complementary slackness conditions fully close the duality gap. However, due to applying the QP duality theory to deal with a quadratic objective function of the lower-level (2.1) to derive the dual, as explained in the Part I paper, Section 2.B, constraint \eqref{eq:PD.1} is also required to obtain zero duality gap. The resulting formulation is nonconvex quadratic.

\begin{equation}\tag{4.10}
\xjedan^2+\xdva^2+\xtri^2 + ... \le \xnula^2
\label{eq:CC.1}
\end{equation}

\begin{equation}\tag{4.11}
\yjedan^2+\ydva^2+\ytri^2 + ... \le \ynula^2
\label{eq:CC.2}
\end{equation}

\begin{equation}\tag{4.12}
\xnula \cdot \ynula + \xjedan \cdot \yjedan + \xdva \cdot \ydva + \xtri \cdot \ytri + ... = 0
\label{eq:CC.3}
\end{equation}

The problem maximizes \eqref{eq:ofqp} subject to constraints (1.2)--(1.7), (2.2)--(2.14), excluding (2.8.1) and (2.9.1) in favor of (3.1.1)--(3.2.4), (B.2)--(B.13), \eqref{eq:PD.1} and constraints based on \eqref{eq:CC.3} (one for every primal-dual inequality pair), with respect to the variables set $\Xiul \cup \Xir \cup \Xidu$.

\subsubsection{Relaxed Complementary Slackness (CS-R)}\hfill

The relaxed complementary slackness technique enhances numerical tractability by allowing a small deviance of the complementary slackness conditions, as in constraint \eqref{eq:CC.4} and shown in Section 12.3.1.1 in \cite{knjiga}. The constraint is only bounded from the upper side since the left-hand side is always nonnegative due to SOC constraints from the primal and the dual. The resulting formulation is nonconvex quadratic.

\begin{equation}\tag{4.13}
\xnula \cdot \ynula + \xjedan \cdot \yjedan + \xdva \cdot \ydva + \xtri \cdot \ytri + ... \le \eps
\label{eq:CC.4}
\end{equation}

The problem maximizes \eqref{eq:ofqp} subject to constraints (1.2)--(1.7), (2.2)--(2.14), excluding (2.8.1) and (2.9.1) in favor of (3.1.1)--(3.2.4), (B.2)--(B.13), \eqref{eq:PD.1} and constraints based on \eqref{eq:CC.4} (one for every primal-dual inequality pair), with respect to the variables set $\Xiul \cup \Xir \cup \Xidu$.

\subsubsection{Aggregated Complementary Slackness (CS-A)}\hfill

Since $x^{\intercal}y$ is always nonnegative due to the primal and dual SOC constraints, complementary conditions can also be aggregated into a single large constraint as in \eqref{eq:CC.5} which sums over all primal and dual vector variable pairs $x$ and $y$ from set $\xi$. The resulting formulation is also nonconvex quadratic.

\begin{equation}\tag{4.14}
 \sum_{(x,y)\in \xi} \!\!\! x^{\intercal}y = 0
\label{eq:CC.5}
\end{equation}

The problem maximizes \eqref{eq:ofqp} subject to constraints (1.2)--(1.7), (2.2)--(2.14), excluding (2.8.1) and (2.9.1) in favor of (3.1.1)--(3.2.4), (B.2)--(B.13), \eqref{eq:PD.1} and the constraint based on \eqref{eq:CC.5}, with respect to the variables set $\Xiul \cup \Xir \cup \Xidu$.

\subsubsection{Relaxed Aggregated Complementary Slackness (CS-AR)}

\indent \indent The same way the individual complementary slackness conditions can be relaxed to potentially improve numerical tractability, the aggregated constraint can be relaxed as well. To make an easier comparison to the nonaggregated version, $\eps$ parameter is enlarged for every primal-dual inequality pair. The resulting formulation is nonconvex quadratic.

\begin{equation}\tag{4.15}
\sum_{(x,y)\in \xi} \!\!\! x^{\intercal}y \le \sum_{(x,y)\in \xi} \!\!\! \eps
\label{eq:CC.6}
\end{equation}


The problem maximizes \eqref{eq:ofqp} subject to constraints (1.2)--(1.7), (2.2)--(2.14), excluding (2.8.1) and (2.9.1) in favor of (3.1.1)--(3.2.4), (B.2)--(B.13), \eqref{eq:PD.1} and constraint based on \eqref{eq:CC.6}, with respect to the variables set $\Xiul \cup \Xir \cup \Xidu$.

\subsection{Penalty Factor}\label{sub:PF}

\subsubsection{Penalty Factor -- Strong Duality (PF-SD)}\hfill

The penalty factor technique closes the duality gap by penalizing it in the main objective function \eqref{eq:PF.1}. Conceptually, as the penalty factor $\pen$ goes to infinity, the duality gap closes to zero. However, the formulation is nonconvex due to a bilinear term in the objective function and thus the global optimality can not be guaranteed and numerical issues may occur for high penalty factors. Properties of the penalty factor techniques are discused in \cite{pen}.

\begin{equation}\tag{4.16}
\mathrm{Max}\:\: \!\sum_{t,i\in \BB} \!\qbat \!\cdot \lambdaJedan +\!\! \sum_{t,i\in \BB} \!\qqbat \!\cdot \lambdaDva + (1+\pen)\cdot(\Omegad - \Omegap)
\label{eq:PF.1}
\end{equation}

The problem maximizes \eqref{eq:PF.1} subject to constraints (1.2)--(1.7), (2.2)--(2.14), excluding (2.8.1) and (2.9.1) in favor of (3.1.1)--(3.2.4) and (B.2)--(B.13), with respect to the variables set $\Xiul \cup \Xir \cup \Xidu$.

\subsubsection{Penalty Factor -- Complementary Slackness (PF-CS)}

\indent \indent Duality gap can also be closed by penalizing the deviance of complementary slackness from zero. The same as the strong duality version, this version is also nonconvex.

\begin{equation}\tag{4.17}
\mathrm{Max}\:\: \!\!\sum_{t,i\in \BB} \!\!\qbat \!\cdot \!\lambdaJedan \!+\!\!\!\! \sum_{t,i\in \BB} \!\!\qqbat \!\cdot\! \lambdaDva \!+\! (\Omegad \!- \Omegap) \!-\! \pen \!\cdot \! \hspace{-7pt}\sum_{(x,y)\in \xi} \!\!\!\! x^{\intercal}y\!
\label{eq:PF.2}
\end{equation}

The problem maximizes \eqref{eq:PF.2} subject to constraints (1.2)--(1.7), (2.2)--(2.14), excluding (2.8.1) and (2.9.1) in favor of (3.1.1)--(3.2.4), (B.2)--(B.13) and \eqref{eq:PD.1}, with respect to the variables set $\Xiul \cup \Xir \cup \Xidu$.

\subsection{Interaction Discretization}

For this group of techniques we define additional sets. $\BSS$ is a set of binary variables for binary expansion variables indexed with letter $\ib$ (i.e. \{1,2,...,$\ceil{\log_2{D}}$\}), and $\US$ is a set of binary variables for unary expansion indexed by $\iu$ (i.e. \{1,2,...,$D$\}), where $D$ is a number of discretization steps. This technique discretizeses the allowed charging $\qch$ and discharging $\qdis$ quantity values so that the bilinear term $\qbat \cdot \lambdaJedan$ from the strong duality can be reformulated into a mixed-integer linear one. Analogous discretizations also need to be done for reactive power. Thus, we introduce variables for consumed and produced reactive powers $\qqch$ and $\qqdis$, respectively, and the binary variable $\xxch$ that prevents simultaneous power exchange in both directions. Their values are defined in constraints \eqref{eq:BE.01}--\eqref{eq:BE.03} in analogous way as in constraints (1.4)--(1.6). The technique can use either the strong duality constraint (in the inequality form) or the complementary-slackness-penalized objective function. Since interaction discretization techniques linearize the bilinear terms, which are the only source of nonconvexity except for the introduced discrete variables, the formulations are of mixed-integer SOCP (MISOCP) class. Techniques belonging to this group can theoretically find proven optimal solution with zero duality gap (assuming high penalty factor for penalty version). General formulation of the expansions can be found in \cite{expansion}. The following constraints split the positive and negative ES reactive power in two variables since they are used separately in the following subsections.

\begin{equation}\tag{4.18}
0 \le \qqch \le \qchmax \cdot \xxch, \quad \forall{t}
\label{eq:BE.01}
\end{equation}

\begin{equation}\tag{4.19}
0 \le \qqdis \le \qdismax \cdot (1-\xxch), \quad \forall{t}
\label{eq:BE.02}
\end{equation}

\begin{equation}\tag{4.20}
\qqbat = \qqch - \qqdis, \quad \forall{t}
\label{eq:BE.03}
\end{equation}

\subsubsection{Binary Expansion -- Strong Duality (BE-SD)}\hfil

Discretization by binary expansion uses binary variables with exponentially increasing (base 2) assigned weights to denote an integer number. The obtained integer number divided by a maximum achievable integer number represents the value the continuous variable will take. If the ratio is 1, the continuous variable takes the upper bound value, and for ratio 0 it takes the lower bound value. The other discrete states are spread evenly. The number of binary variables grows logarithmically with the number of discrete states. The introduced variables are $\ybat,\yybat$, integer variables, $\xbin,\xxbin$ binary variables with assigned weights, and auxiliary variables $\wbe,\wwbe\,\wch,\wdis,\wwch,\wwdis$. Using this technique, the strong duality constraint \eqref{eq:MC.6} takes a convex form since bilinear terms cancel out each other and are effectively replaced by terms $\wch,\wdis,\wwch,\wwdis$.

\begin{equation}\tag{4.21}
\Omegap \!\!\le\! \Omegad \!\!+ \!\!\!  \sum_{t,i\in \BB} \hspace{-3pt} (\qbat \!\cdot \! \lambdaJedan \!+ \! \qqbat \!\cdot \! \lambdaDva) - \!\! \sum_{t}\hspace{-0.5pt}(\wch\!\!\hspace{-0.5pt}-\wdis\!\!+\wwch\!\!\hspace{-0.5pt}-\wwdis)
\label{eq:MC.6}
\end{equation}

\begin{equation}\tag{4.22}
0 \le \begin{pmatrix} \ybat & \!\!\!\! \yybat \end{pmatrix} \le 2^{|\BSS|}-1,\quad \forall{t}
\label{eq:BE.1}
\end{equation}

\begin{equation}\tag{4.23}
\begin{pmatrix} \ybat & \!\!\!\! \yybat \end{pmatrix} = \sum_{\ib \in \BSS}2^{b-1}\cdot \begin{pmatrix}\xbin & \!\!\!\! \xxbin \end{pmatrix},\quad \forall{t}
\label{eq:BE.2}
\end{equation}

\begin{equation}\tag{4.24}
\begin{pmatrix}\qch & \!\!\!\!\qqch\end{pmatrix}\!\!/\qchmax \!+\! \begin{pmatrix}\qdis & \!\!\!\! \qqdis\end{pmatrix}\!\!/\qdismax \!\!=\! \begin{pmatrix}\ybat & \!\!\!\!\yybat\end{pmatrix}\!\!/(2^{|\BSS|}\!-\!1),\quad \!\!\!\!\forall{t}
\label{eq:BE.3}
\end{equation}

\begin{equation}\tag{4.25}
\begin{split}
&\begin{pmatrix}\wch & \!\!\!\!\wwch\end{pmatrix}\!\!/\qchmax \!+\! \begin{pmatrix}\wdis & \!\!\!\!\wwdis\end{pmatrix}\!\!/\qdismax \!= \\
&\sum_{\ib \in \BSS} \! 2^{b-1} \!\!\cdot \! \begin{pmatrix}\wbe & \!\!\!\! \wwbe\end{pmatrix}\!\!/(2^{|\BSS|}-1),\quad \forall{t}
\label{eq:BE.4}
\end{split}
\end{equation}

\begin{equation}\tag{4.26}
\begin{split}
&\begin{pmatrix}\lambdaJedanDn &\!\!\!\!\!  \lambdaDvaDn \end{pmatrix}^{\intercal}  \!\!\!\cdot \! \begin{pmatrix} \xch & \!\!\!\!\! \xxch \end{pmatrix} \! \cdot\! \qchmax \!\! \le \!\! \begin{pmatrix}\wch & \!\!\!\! \wwch \end{pmatrix} \!\!\le \\
&\begin{pmatrix}\lambdaJedanUp &\!\!\!\!\!  \lambdaDvaUp \end{pmatrix}^{\intercal} \!\!\!\cdot \! \begin{pmatrix} \xch & \!\!\!\!\! \xxch \end{pmatrix}\! \cdot\! \qchmax\!,\quad \forall{t}
\label{eq:BE.5}
\end{split}
\end{equation}

\begin{equation}\tag{4.27}
\begin{split}
&\begin{pmatrix}\lambdaJedanDn &\!\!\!\!  \lambdaDvaDn \end{pmatrix}^{\intercal}  \!\!\!\cdot \! \begin{pmatrix} 1-\xch & \!\!\!\! 1-\xxch \end{pmatrix} \! \cdot\! \qchmax \!\! \le \!\! \begin{pmatrix} \wdis & \!\!\!\! \wwdis \end{pmatrix}^{\intercal} \!\!\le \\
&\begin{pmatrix}\lambdaJedanUp &\!\!\!\!  \lambdaDvaUp \end{pmatrix}^{\intercal} \!\!\!\cdot \! \begin{pmatrix} 1-\xch & \!\!\!\! 1-\xxch \end{pmatrix}\! \cdot\! \qchmax\!,\quad \quad \forall{t}\\[-0.95cm]
\label{eq:BE.6}
\end{split}
\end{equation}

\begin{equation}\tag{4.28}
\begin{split}
&\begin{pmatrix}\xbin & \!\!\!\! \xxbin \end{pmatrix}^{\intercal} \!\!\cdot  \begin{pmatrix}\lambdaJedanDn & \!\!\!\! \lambdaDvaDn\end{pmatrix}\!\le \begin{pmatrix}\wbe & \!\!\!\! \wwbe \end{pmatrix}^{\intercal} \! \le \\
&\begin{pmatrix}\xbin & \!\!\!\! \xxbin \end{pmatrix}^{\intercal} \!\!\cdot \begin{pmatrix}\lambdaJedanUp & \!\!\!\! \lambdaDvaUp\end{pmatrix}, \quad  \forall{t,\!i\!\in\!\! \BB,\!\ib \!\in\! \BSS}
\end{split}
\label{eq:BE.7}
\end{equation}

\begin{equation}\tag{4.29}
\begin{split}
&\begin{pmatrix} \lambdaJedan & \hspace{-1pt} \!\!\!\! \lambdaDva \!  \end{pmatrix}^{\!\intercal} \!\!\!+\!\! \hspace{-0.5pt} \begin{pmatrix}\! \xbin & \!\!\!\!\! \xxbin \!\! \end{pmatrix}^{\!\! \intercal} \!\!\!  \cdot \!  \begin{pmatrix} \hspace{-0.8pt} \lambdaJedanUp & \!\!\!\!\! \lambdaDvaUp \! \end{pmatrix} \!\!-\!\hspace{-1pt} \hspace{-1pt} \begin{pmatrix} \hspace{-0.8pt} \lambdaJedanUp & \!\!\!\!\! \lambdaDvaUp \end{pmatrix}^{\!\hspace{-0.5pt}\intercal} \!\!\! \le \\
&\begin{pmatrix}\wbe & \!\!\!\!\!\! \wwbe \end{pmatrix}^{\intercal} \!\! \le \!\!\begin{pmatrix} \lambdaJedan & \!\!\!\!\! \lambdaDva \end{pmatrix}^{\!\intercal} \!\!\!+\!\! \begin{pmatrix} \xbin & \!\!\!\!\!\! \xxbin \end{pmatrix}^{\intercal} \!\!\!\! \cdot \! \hspace{-1pt} \begin{pmatrix}\lambdaJedanDn & \!\!\!\!\! \lambdaDvaDn \end{pmatrix} \\
&-\begin{pmatrix}\lambdaJedanDn & \!\!\!\!\! \lambdaDvaDn \end{pmatrix}^{\intercal}\!\!\!\!, \quad  \forall{t,\!i\!\in\!\! \BB,\!\ib \!\in\! \BSS}\\[-0.95cm]
\label{eq:BE.8}
\end{split}
\end{equation}

The problem maximizes \eqref{eq:ofqp} subject to constraints (1.2)--(1.7), (2.2)--(2.14), excluding (2.8.1) and (2.9.1) in favor of (3.1.1)--(3.2.4), (B.2)--(B.13), \eqref{eq:BE.01}--\eqref{eq:BE.8}, with respect to the variables set $\Xiul \cup \Xir \cup \Xidu \cup \{\wch\!,\wwch\!,\wdis\!,\wwdis\!,\wbe\!,\wwbe\!,\ybat,\yybat,\xbin\!,\xxbin\}$.

\subsubsection{Binary Expansion -- Penalty Factor (BE-PF)}\hfill

The binary expansion technique can also be applied to the bilinear term $\qbat \!\cdot\! \lambdaJedan$ and its reactive power counterpart appearing in the objective function which penalizes the duality gap \eqref{eq:PF.1}. In the resulting objective function \eqref{eq:BE.9}, the bilinear terms from $\Omegad \!- \Omegap$ are first canceled out with an explicit addition of itself and then replaced with their equivalent linear expression $\wch\!-\wdis+\wwch-\wdis$. Except for the lack of the strong duality constraint, the other constraints are the same as for the strong duality version of this technique.

\begin{equation}\tag{4.30}
\begin{split}
&\mathrm{Max}\:\:\!\! \sum_{t,i\in \BB}\!\! (1+\pen) \!\cdot\! (\qbat \!\cdot \! \lambdaJedan \!+\!\qqbat \!\cdot\! \lambdaDva) \!+\! (1\!+\!\pen) \!\cdot\! (\Omegad \!- \Omegap) \\
&- \pen \cdot\! \sum_{t}(\wch\!-\wdis\!+\wwch\!-\wdis)\\[-1.2cm]
\label{eq:BE.9}
\end{split}
\end{equation}

The problem maximizes \eqref{eq:BE.9} subject to constraints (1.2)--(1.7), (2.2)--(2.14), excluding (2.8.1) and (2.9.1) in favor of (3.1.1)--(3.2.4), (B.2)--(B.13), \eqref{eq:BE.01}--\eqref{eq:BE.03} and \eqref{eq:BE.1}--\eqref{eq:BE.8}, with respect to the variables set $\Xiul \cup \Xir \cup \Xidu \cup \{\wch\!,\wwch\!,\wdis\!,\wwdis\!,\wbe\!,\wwbe\!,\ybat,\yybat,\xbin\!,\xxbin\}$.

\subsubsection{Unary Expansion -- Strong Duality (UE-SD)}\hfill

Similarly to the binary expansion, the unary expansion uses binary variables to represent an integer number. However, the assigned weights in this technique increase linearly and only up to one binary variable is allowed to take value 1. The number of binary variables grows linearly with the number of discrete states. In addition to the variables from the binary expansion, the introduced variables are binary variables $\xun,\xxun$ and auxiliary variables $\wue,\wwue$.

\begin{equation}\tag{4.31}
0 \le \begin{pmatrix}\ybat & \!\!\!\! \yybat \end{pmatrix} \le |\US|,\quad \forall{t}
\label{eq:UE.1}
\end{equation}

\begin{equation}\tag{4.32}
\begin{pmatrix} \ybat & \!\!\!\! \yybat \end{pmatrix} = \sum_{\iu\in\US}\iu\cdot \begin{pmatrix} \xun & \!\!\!\! \xxun \end{pmatrix},\quad \forall{t}
\label{eq:UE.2}
\end{equation}

\begin{equation}\tag{4.33}
\sum_{\iu \in \US} \begin{pmatrix}\xun & \!\!\!\! \xxun \end{pmatrix} \le 1,\quad \forall{t}
\label{eq:UE.3}
\end{equation}

\begin{equation}\tag{4.34}
\begin{pmatrix}\qch & \!\!\!\! \qqch \end{pmatrix}\!\!/\qchmax + \begin{pmatrix}\qdis & \!\!\!\! \qqdis \end{pmatrix}\!\!/\qdismax = \begin{pmatrix}\ybat & \!\!\!\! \yybat \end{pmatrix}\!\!/|\US|,\quad \forall{t}
\label{eq:UE.4}
\end{equation}

\begin{equation}\tag{4.35}
\begin{pmatrix}\wch & \!\!\!\! \wwch \end{pmatrix}\!\!/\qchmax \!+\! \begin{pmatrix}\wdis & \!\!\!\! \wwdis \end{pmatrix}\!\!/\qdismax \!= \!\!\!\sum_{\iu \in \US} \!\!\iu \cdot \wue / |\US|,\quad \!\!\! \forall{t}
\label{eq:UE.5}
\end{equation}

\begin{equation}\tag{4.36}
\begin{split}
&\begin{pmatrix}\xun & \!\!\!\! \xxun \end{pmatrix}^{\intercal} \!\!\cdot  \begin{pmatrix}\lambdaJedanDn & \!\!\!\! \lambdaDvaDn\end{pmatrix}\!\le \begin{pmatrix}\wue & \!\!\!\! \wwue \end{pmatrix}^{\intercal} \! \le \\
&\begin{pmatrix}\xun & \!\!\!\! \xxun \end{pmatrix}^{\intercal} \!\!\cdot \begin{pmatrix}\lambdaJedanUp & \!\!\!\! \lambdaDvaUp\end{pmatrix}, \quad  \forall{t,\!i\!\in\!\! \BB,\!\ib \!\in\! \BSS}
\end{split}
\label{eq:UE.6}
\end{equation}

\begin{equation}\tag{4.37}
\begin{split}
&\begin{pmatrix} \lambdaJedan & \hspace{-1pt} \!\!\!\! \lambdaDva \!  \end{pmatrix}^{\!\intercal} \!\!\!+\!\!  \begin{pmatrix} \xun & \!\!\!\!\! \xxun \!\! \end{pmatrix}^{\!\! \intercal} \!\!\!  \cdot \!  \begin{pmatrix} \hspace{-0.8pt} \lambdaJedanUp & \!\!\!\!\! \lambdaDvaUp \! \end{pmatrix} \!\!-\!\hspace{-1pt} \hspace{-1pt} \begin{pmatrix} \hspace{-0.8pt} \lambdaJedanUp & \!\!\!\!\! \lambdaDvaUp \end{pmatrix}^{\!\hspace{-0.5pt}\intercal} \!\!\! \le \\
&\begin{pmatrix}\wue & \!\!\!\!\!\! \wwue \end{pmatrix}^{\intercal} \!\! \le \!\!\begin{pmatrix} \lambdaJedan & \!\!\!\!\! \lambdaDva \end{pmatrix}^{\!\intercal} \!\!\!+\!\! \begin{pmatrix} \xun & \!\!\!\!\!\! \xxun \end{pmatrix}^{\intercal} \!\!\!\! \cdot \! \hspace{-1pt} \begin{pmatrix}\lambdaJedanDn & \!\!\!\!\! \lambdaDvaDn \end{pmatrix} \\
&-\begin{pmatrix}\lambdaJedanDn & \!\!\!\!\! \lambdaDvaDn \end{pmatrix}^{\intercal}\!\!\!\!, \quad  \forall{t,\!i\!\in\!\! \BB,\!\ib \!\in\! \BSS}\\[-0.95cm]
\label{eq:UE.7}
\end{split}
\end{equation}

The problem maximizes \eqref{eq:ofqp} subject to constraints (1.2)--(1.7), (2.2)--(2.14) excluding (2.8.1) and (2.9.1) in favor of (3.1.1)--(3.2.4), (B.2)--(B.13), \eqref{eq:BE.01}--\eqref{eq:MC.6}, \eqref{eq:BE.5}, \eqref{eq:BE.6} and \eqref{eq:UE.1}--\eqref{eq:UE.7} with respect to the variables set $\Xiul \cup \Xir \cup \Xidu \cup \{\wch,\wwch\wdis,\wwdis,\wue,\wwue,\ybat,\yybat,\xun,\xxun\}$.

\subsubsection{Unary Expansion -- Penalty Factor (UE-PF)}\hfill

The same as the binary expansion, the unary expansion can also be applied to the penalized objective function \eqref{eq:BE.9}. Other constraints, except for the lack of the strong duality constraint, are the same as for the strong duality version of the unary expansion.

The problem maximizes \eqref{eq:BE.9} subject to constraints (1.2)--(1.7), (2.2)--(2.14), excluding (2.8.1) and (2.9.1) in favor of (3.1.1)--(3.2.4), (B.2)--(B.13), \eqref{eq:BE.01}--\eqref{eq:BE.03}, \eqref{eq:BE.5}, \eqref{eq:BE.6} and \eqref{eq:UE.1}--\eqref{eq:UE.7}, with respect to the variables set $\Xiul \cup \Xir \cup \Xidu \cup \{\wch,\wwch,\wdis,\wwdis,\wue,\wwue,\ybat,\yybat,\xun,\xxun\}$.


\subsection{Smoothing techniques}\label{sub:SM}
In this section we consider smoothing of the complementary slackness conditions. Smoothing techniques are presented in \cite{sm}, where they are generalized from smoothing the linear complementary conditions to smoothing the SOCP complementary conditions. The resulting equality constraints replace both the primal-dual inequality pair and the complementary slackness conditions \eqref{eq:CC.1}--\eqref{eq:CC.3} for SOC and linear inequalities. The only inequality constraints left in the model are those from the upper-level problem. Smoothing techniques become exact as $\eps$ parameter approaches zero from the positive side. Formulations using the smoothing techniques belong to the nonconvex nonlinear class.

\subsubsection{Chen–Harker–Kanzow–Smale (SM1)}\hfill
\label{subsub:sm1}

This complementary slackness conditions smoothing technique is a special case of the Chen-Mangasarian smoothing functions. Only the vector constraint \eqref{eq:SM1.1} is directly part of the optimization problem. Formulas \eqref{eq:SM1.2}--\eqref{eq:SM1.4} only define parts for substitution of constraint \eqref{eq:SM1.1}. Specifically, \eqref{eq:SM1.2} defines a function, \eqref{eq:SM1.3} defines a numerical expression and \eqref{eq:SM1.4} defines a vector.

\begin{equation}\tag{4.38}
\x - \eps \cdot \left(\smf(\psijedan/\eps)\cdot \ujedan  + \smf(\psidva/\eps)\cdot \udva \right)= 0
\label{eq:SM1.1}
\end{equation}

\begin{equation}\tag{4.39}
\smf (\alpha) = (\sqrt{\alpha^2 + 4} + \alpha)/2
\label{eq:SM1.2}
\end{equation}

\begin{equation}\tag{4.40}
\psin = \xnula - \ynula + (-1)^n \cdot \norm{\xcrta-\ycrta}
\label{eq:SM1.3}
\end{equation}

\begin{equation}\tag{4.41}
\un = \begin{pmatrix} 1/2 & 1/2\cdot (-1)^n \cdot \frac{\xcrta-\ycrta}{\norm{\xcrta-\ycrta}} \end{pmatrix}
\label{eq:SM1.4}
\end{equation}

The problem maximizes \eqref{eq:ofqp} subject to constraints (1.2)--(1.7), (2.2)--(2.7), (2.8.2), (2.9.2), (2.13), (3.1.2)--(3.1.5), (3.2.2)--(3.2.4), (B.2)--(B.9), \eqref{eq:PD.1} and constraints based on \eqref{eq:SM1.1} (one vector constraint for every primal-dual inequality and complementary condition pair), whose parts are defined in \eqref{eq:SM1.2}--\eqref{eq:SM1.4}, with respect to the variables set $\Xiul \cup \Xir \cup \Xidu$.

\subsubsection{Kanzow (SM2)}\hfill

This smoothing function is a variation of the Fischer-Burmeister function. Analogous to the previous smoothing technique, \eqref{eq:SM2.1} is a vector constraint, while \eqref{eq:SM2.2} and \eqref{eq:SM2.3} are only used to define parts of \eqref{eq:SM2.1}.

\begin{equation}\tag{4.42}
\x + \y - (\sqrt{\psijedan}\cdot \ujedan + \sqrt{\psidva}\cdot \udva)=0
\label{eq:SM2.1}
\end{equation}

\begin{equation}\tag{4.43}
\psin = \norm{\x}^2 + \norm{\y}^2 + 2 \!\cdot\! \eps^2 + 2\!\cdot\!(-1)^n \!\cdot\! \norm{\xnula \!\cdot\! \xcrta + \ynula\!\cdot\! \ycrta}
\label{eq:SM2.2}
\end{equation}

\begin{equation}\tag{4.44}
\un = \begin{pmatrix} 1/2  &  1/2\!\cdot\! (-1)^n\!\cdot\!\frac{\xnula \!\cdot\! \xcrta + \ynula\!\cdot\! \ycrta}{\norm{\xnula \!\cdot\! \xcrta + \ynula\!\cdot\! \ycrta}}\end{pmatrix}
\label{eq:SM2.3}
\end{equation}

The problem maximizes \eqref{eq:ofqp} subject to constraints (1.2)--(1.7), (2.2)--(2.7), (2.8.2), (2.9.2), (2.13), (3.1.2)--(3.1.5), (3.2.2)--(3.2.4), (B.2)--(B.9), \eqref{eq:PD.1} and constraints based on \eqref{eq:SM2.1} (one vector constraint for every primal-dual inequality and complementary condition pair), whose parts are defined in \eqref{eq:SM2.2} and \eqref{eq:SM2.3}, with respect to the variables set $\Xiul \cup \Xir \cup \Xidu$.

\begin{table*}[t]
    \setlength{\tabcolsep}{0.86pt} 
    \renewcommand{\arraystretch}{1} 
    \caption{Model accuracy tested on 3\_lmbd network (IP -- interior point, S -- simplex)}
\label{tab:net3}
\centering
\scriptsize
\begin{tabular}{lcccccccccccccccccc|}
\cline{2-19}
\multicolumn{1}{l|}{}                                     & \multicolumn{6}{c|}{ES at bus 1}                                                                                                                                                                                         & \multicolumn{6}{c|}{ES at bus 2}                                                                                                                                                                                         & \multicolumn{6}{c|}{ES at bus 3}                                                                                                                                                                    \\ \cline{2-19} 
\multicolumn{1}{l|}{}                                     & \multicolumn{3}{c|}{ES profit}                                                                             & \multicolumn{3}{c|}{System expenses}                                                                        & \multicolumn{3}{c|}{ES profit}                                                                             & \multicolumn{3}{c|}{System expenses}                                                                        & \multicolumn{3}{c|}{ES profit}                                                                             & \multicolumn{3}{c|}{System expenses}                                                   \\ \hline
\multicolumn{19}{|c|}{ES active-power-only bids}                                                                                                                                                                                                                                                                                                                                                                                                                                                                                                                                                                                                                                                                      \\ \hline
\multicolumn{1}{|l|}{Model}                               & Actual       & Computed     & \multicolumn{1}{c|}{\begin{tabular}[c]{@{}c@{}}Diff\\ {[}\%{]}\end{tabular}} & Actual        & Computed     & \multicolumn{1}{c|}{\begin{tabular}[c]{@{}c@{}}Diff\\ {[}\%{]}\end{tabular}} & Actual       & Computed     & \multicolumn{1}{c|}{\begin{tabular}[c]{@{}c@{}}Diff\\ {[}\%{]}\end{tabular}} & Actual        & Computed     & \multicolumn{1}{c|}{\begin{tabular}[c]{@{}c@{}}Diff\\ {[}\%{]}\end{tabular}} & Actual       & Computed     & \multicolumn{1}{c|}{\begin{tabular}[c]{@{}c@{}}Diff\\ {[}\%{]}\end{tabular}} & Actual        & Computed     & \begin{tabular}[c]{@{}c@{}}Diff\\ {[}\%{]}\end{tabular} \\ \hline
\multicolumn{1}{|l|}{CPSOTA}                              & 1818.65      & 1818.56      & \multicolumn{1}{c|}{-4.9e-3}                                                    & 98827.27      & 98827.34     & \multicolumn{1}{c|}{7.1e-5}                                                    & 1359.88      & 1359.80      & \multicolumn{1}{c|}{-5.9e-3}                                                   & 99342.16      & 99342.20     & \multicolumn{1}{c|}{4.0e-5}                                                    & 2016.85      & 2017.79      & \multicolumn{1}{c|}{0.047}                                                    & 98496.77      & 98496.94     & 1.7e-4                                                    \\ \hline
\multicolumn{1}{|l|}{Jabr's \cite{jabr}} & 1817.61      & 1641.17      & \multicolumn{1}{c|}{-9.7}                                                   & 98821.80      & 98443.00     & \multicolumn{1}{c|}{-0.38}                                                   & 1327.67      & 1582.75      & \multicolumn{1}{c|}{19.2}                                                   & 99327.24      & 98506.39     & \multicolumn{1}{c|}{-0.83}                                                   & 2007.06      & 1754.30      & \multicolumn{1}{c|}{-13}                                                  & 98486.16      & 98304.91     & -0.18                                                   \\ \hline
\multicolumn{1}{|l|}{DC}                                  & 1818.35      & 1757.64      & \multicolumn{1}{c|}{-3.3}                                                   & 98824.47      & 97319.61     & \multicolumn{1}{c|}{-1.5}                                                   & 1359.04      & 1364.71      & \multicolumn{1}{c|}{0.42}                                                    & 99341.82      & 97774.13     & \multicolumn{1}{c|}{-1.6}                                                   & 1986.50      & 1729.57      & \multicolumn{1}{c|}{-13}                                                  & 98531.12      & 97236.05     & -1.3                                                   \\ \hline
\multicolumn{1}{|l|}{Centralized}                         &  \multicolumn{2}{c}{1795.61} & \multicolumn{1}{c|}{}                                                        & \multicolumn{2}{c}{98811.38} & \multicolumn{1}{c|}{}                                                        & \multicolumn{2}{c}{1347.65} & \multicolumn{1}{c|}{}                                                        & \multicolumn{2}{c}{99317.95} & \multicolumn{1}{c|}{}                                                        & \multicolumn{2}{c}{1988.96} & \multicolumn{1}{c|}{}                                                        & \multicolumn{2}{c}{98481.34} &                                                         \\ \hline
\multicolumn{1}{|l|}{Fixed prices -- IP}                   & 1329.03      & 2224.99      & \multicolumn{1}{c|}{67}                                                   & 99013.29      & 98529.70     & \multicolumn{1}{c|}{-0.49}                                                   & 719.18       & 1560.60      & \multicolumn{1}{c|}{117}                                                  & 99614.43      & 99194.09     & \multicolumn{1}{c|}{-0.42}                                                   & 1138.13      & 2915.78      & \multicolumn{1}{c|}{156}                                                  & 98966.00      & 97838.91     & -1.1                                                   \\ \hline
\multicolumn{1}{|l|}{Fixed prices -- S}                    & 1122.16      & 2224.99      & \multicolumn{1}{c|}{98}                                                   & 99122.00      & 98529.70     & \multicolumn{1}{c|}{-0.60}                                                   & 490.31       & 1560.60      & \multicolumn{1}{c|}{218}                                                  & 99729.13      & 99194.09     & \multicolumn{1}{c|}{-0.54}                                                   & -150.98      & 2915.78      & \multicolumn{1}{c|}{-2031}                                                & 99345.15      & 97838.91     & -1.5                                                   \\ \hline
\end{tabular}
\end{table*}

\section{Case Study}\label{sec:case}

\subsection{Description and Set-Up}
\label{sub:description}
The case study consists of three parts. The first one demonstrates the accuracy of our convex polar second-order Taylor approximation (CPSOTA) model \cite{cpsota} in comparison to implementations that use the existing convex OPF formulations in the lower level, i.e. Jabr and DC. The accuracy comparison is based on the smoothing solution technique that effectively ensures strong duality ($\Omegap\!=\!\Omegad$). The second case study shows economic benefits for the ES and the system by comparing profits with only active power bids from the ES and both reactive and active power bids. The third case study evaluates different solution techniques with focus on accuracy, i.e. duality gap and objective function value, and numerical tractability.

The case study considers a large energy storage unit with 100 MWh capacity (1 p.u.), 60 MW (0.6 p.u.) maximum (dis)charging rate and 0.9 (dis)charging efficiency. The initial state-of-energy is set to 50\% and the storage is allowed to end a day at any state-of-energy. All variable and parameter units are in p.u. or dimensionless.

Transmission system meshed networks were taken from the PGLib-OPF v19.05 \cite{pglib} database. A 24-hour time horizon was added by scaling the loads with winter weekday profile factors from IEEE RTS-96 \cite{rts96}. 

All problems were solved on a desktop PC (i7 9700; 32 GB, 2.67 GHz RAM) in AMPL. Convex and mixed-integer convex problems were solved in Xpress 8.10.1, while all other problems in KNITRO 12.3. The default solver settings were used, except for the Algorithm 1 step 5 (see Part I) for which the settings are stated individually in the case studies. Algorithms 1 steps 1--4 and 6 are single-level continuous optimizations and thus easy to solve. Their run times are less than a second even for the largest considered network with 57 buses. The threshold for imposing the line thermal power limits, controlled with Boolean parameter $\conds$, was set to 85\% and no final solution violated the thermal limits.

\subsection{Case Study I: Model Accuracy}
\label{sub:accuracy}

Accuracy of the considered bilevel model mostly depends on the accuracy of the OPF in its lower level and the technique applied to ensure that the strong duality holds at the solution point. Since convexity of OPF is a requirement for strong duality, this case study compares our proposed model with the models based on commonly used convex OPFs: Jabr's \cite{jabr} and DC. In addition, the bilevel approach is compared to two single-level simplifications. One assumes constant prices in the lower level regardless of the energy storage bidding strategy, and the other, centralized, models the ES as if it were owned by the system operator who minimizes the system costs. Prices for the first simplification are obtained in a pre-run by considering an idle energy storage, thus preventing its impact on market prices. For bilevel models, the strong duality is ensured with Chen–Harker–Kanzow–Smale (SM1) smoothing technique, described in Subsection \ref{subsub:sm1}. This technique's defining parameter $\eps$ was set to 1e-3, which in practice ensures a duality gap of 1e-5\% or better. To make the comparison easier, but without the loss of accuracy due to positive prices, i.e. simultaneous charging and discharging does not occur, models were run without the binary variable $\xch$ for energy storage (ES) (dis)charging, otherwise present in constraints (1.4) and (1.5) in the Part I paper. Since the final model (Algorithm 1 step 5) is nonconvex-nonlinear, KNITRO multistart feature (16 starting points, $\pm$0.6 variable perturbations) was used to the increase chances of finding a global optimal solution. Numerical stability was increased by tightening the solver default convergence and infeasibility tolerances by a factor of 100 and by enabling the solver's warm start option. To better differentiate between active and reactive power accuracy, the case study contains separate analysis for the cases when ES can bid only active power and when it can bid both active and reactive power. The exception are the buses with were zero reactive power prices observed at the assumed operating point (Algorithm 1 Step 1) as in that case the two solutions are the same. 

Accuracy comparison for three- and five-bus networks are presented in Tables \ref{tab:net3} and \ref{tab:net5}. The analysis is performed for an ES placed at each bus individually. Table data columns include actual ES profits and system expenses, i.e. verified quantities using Algorithm 1 Step 6, computed profits or expenses, using Algorithm 1 Step 5, and the percentage difference between the actual and the computed profits. The rows include considered models: CPSOTA, Jabr's \cite{jabr}, DC, centralized and fixed-price models solved using two different solver methods (interior point -- IP and simplex -- S). The tree-bus network table does not include active and reactive power bids data since the reactive power prices at all buses are zero. Similarly, the five-bus network table does not include data for active and reactive power bids at buses 4 and 5 due to zero reactive power prices. In the case of zero reactive power prices, the results are the same as in the case of active-power-only bids.


\begin{table*}[h]
    \setlength{\tabcolsep}{0.6pt} 
    \renewcommand{\arraystretch}{1} 
    \caption{Models' accuracy tested on 5\_pjm network (IP -- interior point, S -- simplex)}
\label{tab:net5}
\centering
\scriptsize
\begin{tabular}{llccccccccccccccc|}
\cline{3-17}
                                                                           & \multicolumn{1}{l|}{}                & \multicolumn{3}{c|}{ES at bus 1}                                                                                                             & \multicolumn{3}{c|}{ES at bus 2}                                                                                                             & \multicolumn{3}{c|}{ES at bus 3}                                                                                                             & \multicolumn{3}{c|}{ES at bus 4}                                                                                                             & \multicolumn{3}{c|}{ES at bus 5}                                                                                        \\ \hline
\multicolumn{2}{|l|}{Model}                                                                                       & Actual                        & Computed                      & \multicolumn{1}{c|}{\begin{tabular}[c]{@{}c@{}}Diff\\ {[}\%{]}\end{tabular}} & Actual                        & Computed                      & \multicolumn{1}{c|}{\begin{tabular}[c]{@{}c@{}}Diff\\ {[}\%{]}\end{tabular}} & Actual                        & Computed                      & \multicolumn{1}{c|}{\begin{tabular}[c]{@{}c@{}}Diff\\ {[}\%{]}\end{tabular}} & Actual                        & Computed                      & \multicolumn{1}{c|}{\begin{tabular}[c]{@{}c@{}}Diff\\ {[}\%{]}\end{tabular}} & Actual                        & Computed                      & \begin{tabular}[c]{@{}c@{}}Diff\\ {[}\%{]}\end{tabular} \\ \hline
\multicolumn{17}{|c|}{ES active-power-only bids}                                                                                                                                                                                                                                                                                                                                                                                                                                                                                                                                                                                                                                                                                                                                                                                        \\ \hline
\multicolumn{1}{|l|}{\multirow{2}{*}{CPSOTA}}                              & \multicolumn{1}{l|}{ES profit}       & 804.94                        & 804.94                        & \multicolumn{1}{c|}{ 1.5e-4}                                                    & 1648.09                       & 1648.07                       & \multicolumn{1}{c|}{ -7.7e-4}                                                    & 1958.23                       & 1958.22                       & \multicolumn{1}{c|}{ -5.4e-4}                                                    & 2833.07                       & 2833.10                       & \multicolumn{1}{c|}{ 9.5e-4}                                                    & 696.45                        & 696.38                        &  -0.010                                                   \\
\multicolumn{1}{|l|}{}                                                     & \multicolumn{1}{l|}{System expenses} & \multicolumn{1}{l}{295944.24} & \multicolumn{1}{l}{295944.23} & \multicolumn{1}{c|}{ -1.0e-6}                                                    & \multicolumn{1}{l}{295100.82} & \multicolumn{1}{l}{295100.82} & \multicolumn{1}{c|}{ 1.4e-6}                                                    & \multicolumn{1}{l}{294790.81} & \multicolumn{1}{l}{294790.81} & \multicolumn{1}{c|}{ -2.8e-7}                                                    & \multicolumn{1}{l}{293915.06} & \multicolumn{1}{l}{293915.04} & \multicolumn{1}{c|}{ -6.5e-6}                                                    & \multicolumn{1}{l}{296050.47} & \multicolumn{1}{l}{296050.49} &  -5.9e-6                                                   \\ \hline
\multicolumn{1}{|l|}{\multirow{2}{*}{Jabr's  \cite{jabr}}} & \multicolumn{1}{l|}{ES profit}       & 792.26                        & 1923.73                       & \multicolumn{1}{c|}{ 143}                                                  & 1648.07                       & 1956.98                       & \multicolumn{1}{c|}{ 19}                                                   & 1958.23                       & 1958.22                       & \multicolumn{1}{c|}{ 5.1e-4}                                                    & 2833.06                       & 1950.47                       & \multicolumn{1}{c|}{ -31}                                                  & 171.65                        & 1916.34                       &  1016                                                 \\
\multicolumn{1}{|l|}{}                                                     & \multicolumn{1}{l|}{System expenses} & \multicolumn{1}{l}{295956.89} & \multicolumn{1}{l}{257518.89} & \multicolumn{1}{c|}{ -13}                                                  & \multicolumn{1}{l}{295100.80} & \multicolumn{1}{l}{257485.69} & \multicolumn{1}{c|}{ -13}                                                  & \multicolumn{1}{l}{294790.81} & \multicolumn{1}{l}{257484.66} & \multicolumn{1}{c|}{ -13}                                                  & \multicolumn{1}{l}{293915.01} & \multicolumn{1}{l}{257491.99} & \multicolumn{1}{c|}{ -12}                                                  & \multicolumn{1}{l}{296577.55} & \multicolumn{1}{l}{257526.16} &  -13                                                  \\ \hline
\multicolumn{1}{|l|}{\multirow{2}{*}{DC}}                                  & \multicolumn{1}{l|}{ES profit}       & 755.19                        & 839.01                        & \multicolumn{1}{c|}{ 11}                                                   & 1590.18                       & 1676.82                       & \multicolumn{1}{c|}{ 5.4}                                                    & 1901.61                       & 2002.22                       & \multicolumn{1}{c|}{ 5.3}                                                    & 2761.95                       & 2897.07                       & \multicolumn{1}{c|}{ 4.9}                                                    & 655.15                        & 695.08                        &  6.1                                                    \\
\multicolumn{1}{|l|}{}                                                     & \multicolumn{1}{l|}{System expenses} & \multicolumn{1}{l}{295947.22} & \multicolumn{1}{l}{294248.75} & \multicolumn{1}{c|}{ -0.57}                                                   & \multicolumn{1}{l}{295106.07} & \multicolumn{1}{l}{293410.94} & \multicolumn{1}{c|}{ -0.57}                                                   & \multicolumn{1}{l}{294795.54} & \multicolumn{1}{l}{293085.54} & \multicolumn{1}{c|}{ -0.58}                                                   & \multicolumn{1}{l}{293930.42} & \multicolumn{1}{l}{292190.70} & \multicolumn{1}{c|}{ -0.59}                                                   & \multicolumn{1}{l}{296066.12} & \multicolumn{1}{l}{294392.69} &  -0.57                                                   \\ \hline
\multicolumn{1}{|l|}{\multirow{2}{*}{ Centralized}}                         & \multicolumn{1}{l|}{ ES profit}       & \multicolumn{2}{c}{ 761.87}                                    & \multicolumn{1}{l|}{}                                                        & \multicolumn{2}{c}{ 1601.78}                                   & \multicolumn{1}{l|}{}                                                        & \multicolumn{2}{c}{ 1911.97}                                   & \multicolumn{1}{l|}{}                                                        & \multicolumn{2}{c}{ 2786.38}                                   & \multicolumn{1}{l|}{}                                                        & \multicolumn{2}{c}{ 627.35}                                    & \multicolumn{1}{l|}{}                                   \\
\multicolumn{1}{|l|}{}                                                     & \multicolumn{1}{l|}{ System expenses} & \multicolumn{2}{c}{ 295944.21}                                 & \multicolumn{1}{l|}{}                                                        & \multicolumn{2}{c}{ 295100.67}                                 & \multicolumn{1}{l|}{}                                                        & \multicolumn{2}{c}{ 294790.66}                                 & \multicolumn{1}{l|}{}                                                        & \multicolumn{2}{c}{ 293914.75}                                 & \multicolumn{1}{l|}{}                                                        & \multicolumn{2}{c}{ 296049.57}                                 & \multicolumn{1}{l|}{}                                   \\ \hline
\multicolumn{1}{|l|}{\multirow{2}{*}{ Fixed prices \!--\! IP}}                   & \multicolumn{1}{l|}{ES profit}       & 486.53                        & 1083.59                       & \multicolumn{1}{c|}{ 123}                                                  & 1336.49                       & 1945.05                       & \multicolumn{1}{c|}{ 46}                                                   & 1647.12                       & 2255.19                       & \multicolumn{1}{c|}{ 37}                                                   & 2468.82                       & 3131.36                       & \multicolumn{1}{c|}{ 27}                                                   & -149.81                       & 864.25                        &  -677                                                 \\
\multicolumn{1}{|l|}{}                                                     & \multicolumn{1}{l|}{System expenses} & \multicolumn{1}{l}{296089.10} & \multicolumn{1}{l}{295665.66} & \multicolumn{1}{c|}{ -0.14}                                                   & \multicolumn{1}{l}{295236.83} & \multicolumn{1}{l}{294804.20} & \multicolumn{1}{c|}{ -0.15}                                                   & \multicolumn{1}{l}{294926.79} & \multicolumn{1}{l}{294494.06} & \multicolumn{1}{c|}{ -0.15}                                                   & \multicolumn{1}{l}{294051.53} & \multicolumn{1}{l}{293617.89} & \multicolumn{1}{c|}{ -0.15}                                                   & \multicolumn{1}{l}{296622.61} & \multicolumn{1}{l}{295885.00} &  -0.25                                                   \\ \hline
\multicolumn{1}{|l|}{\multirow{2}{*}{ Fixed prices \!--\! S}}                    & \multicolumn{1}{l|}{ES profit}       & 486.43                        & 1083.59                       & \multicolumn{1}{c|}{ 123}                                                  & 1336.29                       & 1945.05                       & \multicolumn{1}{c|}{ 46}                                                   & 1647.11                       & 2255.19                       & \multicolumn{1}{c|}{ 37}                                                   & 2191.22                       & 3131.36                       & \multicolumn{1}{c|}{ 43}                                                   & -183.87                       & 864.25                        &  -570                                                 \\
\multicolumn{1}{|l|}{}                                                     & \multicolumn{1}{l|}{System expenses} & \multicolumn{1}{l}{296089.15} & \multicolumn{1}{l}{295665.66} & \multicolumn{1}{c|}{ -0.14}                                                   & \multicolumn{1}{l}{295236.93} & \multicolumn{1}{l}{294804.20} & \multicolumn{1}{c|}{ -0.15}                                                   & \multicolumn{1}{l}{294926.79} & \multicolumn{1}{l}{294494.06} & \multicolumn{1}{c|}{ -0.15}                                                   & \multicolumn{1}{l}{294055.85} & \multicolumn{1}{l}{293617.89} & \multicolumn{1}{c|}{ -0.15}                                                   & \multicolumn{1}{l}{296610.75} & \multicolumn{1}{l}{295885.00} &  -0.24                                                   \\ \hline
\multicolumn{17}{|c|}{ ES active and reactive power bids}                                                                                                                                                                                                                                                                                                                                                                                                                                                                                                                                                                                                                                                                                                                                                                                \\ \hline
\multicolumn{1}{|l|}{\multirow{2}{*}{CPSOTA}}                              & \multicolumn{1}{l|}{ES profit}       & 1170.06                       & 1169.34                       & \multicolumn{1}{c|}{ -0.061}                                                   & 1999.66                       & 1989.00                       & \multicolumn{1}{c|}{ -0.53}                                                   & 2016.55                       & 2008.37                       & \multicolumn{1}{c|}{ -0.41}                                                   &  \multicolumn{3}{c|}{\multirow{6}{*}{\diagbox[dir=NW,height=1.2\rotheadsize, width=\dimexpr\eqboxwidth{AB}+138\tabcolsep\relax]%
    {\raisebox{0ex}{}}{\raisebox{-10.4ex}{}}}}                                                                                                      & \multicolumn{3}{c|}{\multirow{6}{*}{\diagbox[dir=NW,height=1.2\rotheadsize, width=\dimexpr\eqboxwidth{AB}+138\tabcolsep\relax]%
    {\raisebox{0ex}{}}{\raisebox{-10.4ex}{}}}}                                                                                 \\
\multicolumn{1}{|l|}{}                                                     & \multicolumn{1}{l|}{System expenses} & 295575.18                     & 295575.52                     & \multicolumn{1}{c|}{ 1.2e-4}                                                    & 294728.23                     & 294733.50                     & \multicolumn{1}{c|}{ 1.8e-3}                                                    & 294711.51                     & 294715.58                     & \multicolumn{1}{c|}{ 1.4e-3}                                                    & \multicolumn{3}{c|}{}                                                                                                                        & \multicolumn{3}{c|}{}                                                                                                   \\ \cline{1-11}
\multicolumn{1}{|l|}{\multirow{2}{*}{Jabr's   \cite{jabr}}} & \multicolumn{1}{l|}{ES profit}       & 951.65                        & 1923.73                       & \multicolumn{1}{c|}{ 102}                                                  & 1982.75                       & 1980.23                       & \multicolumn{1}{c|}{ -0.13}                                                   & 1958.22                       & 1966.43                       & \multicolumn{1}{c|}{ 0.42}                                                    & \multicolumn{3}{c|}{}                                                                                                                        & \multicolumn{3}{c|}{}                                                                                                   \\
\multicolumn{1}{|l|}{}                                                     & \multicolumn{1}{l|}{System expenses} & 295796.91                     & 257518.89                     & \multicolumn{1}{c|}{ -13}                                                  & 294747.12                     & 257462.46                     & \multicolumn{1}{c|}{ -13}                                                  & 294782.46                     & 257484.66                     & \multicolumn{1}{c|}{ -13}                                                  & \multicolumn{3}{c|}{}                                                                                                                        & \multicolumn{3}{c|}{}                                                                                                   \\ \cline{1-11}
\multicolumn{1}{|l|}{\multirow{2}{*}{ Centralized}}                         & \multicolumn{1}{l|}{ ES profit}       & \multicolumn{2}{c}{ 1127.21}                                   & \multicolumn{1}{c|}{}                                                        & \multicolumn{2}{c}{ 1952.71}                                   & \multicolumn{1}{c|}{}                                                        & \multicolumn{2}{c}{ 1970.94}                                   & \multicolumn{1}{c|}{}                                                        & \multicolumn{3}{c|}{}                                                                                                                        & \multicolumn{3}{c|}{}                                                                                                   \\
\multicolumn{1}{|l|}{}                                                     & \multicolumn{1}{l|}{ System expenses} & \multicolumn{2}{c}{ 295575.17}                                 & \multicolumn{1}{c|}{}                                                        & \multicolumn{2}{c}{ 294727.43}                                 & \multicolumn{1}{c|}{}                                                        & \multicolumn{2}{c}{ 294708.75}                                 & \multicolumn{1}{c|}{}                                                        & \multicolumn{3}{c|}{}                                                                                                                        & \multicolumn{3}{c|}{}                                                                                                   \\ \hline
\end{tabular}
\end{table*}

The results in Table \ref{tab:net3} indicate that, despite the significant inaccuracy (actual vs. computed columns), all bilevel approaches still make generally good decisions for an ES bidding active power. For example, for ES at bus 1, Jabr's model underestimates the ES profit at only 1641.17 (10\% underestimated), while the actual profit 1817.61 is very close to the best achieved actual profit of 1818.65. On the other hand, for the ES located at bus 2, the Jabr's model greatly overestimates the computed profit at 1582.75 (16\% overestimated), while the actual profit is only 2.4\% away from the best actual value, achieve by CPSOTA. For the ES at bus 3, the Jabr's model again underestimates the profit. For the DC model, when the ES is located at buses 1 or 2, the computed and actual ES profits are quite close, indicating that the model provides good estimates of the ES profit regardless on the lossless network representation. However, when the ES is at bus 3, which is the bus without generators capable of producing active power and is under the effect of congestion during multiple hours, the DC model underperforms. The centralized model achieves the lowest system expenses in all cases. However, the ES profits are worse then with the DC model at first two buses and better at the third. The fixed price approaches are inadequate for the considered systems. The actual profits are unfavourable and much lower than the computed ones. These severe differences between the computed and the actual profits when neglecting the impact of being a strategic player is in line with findings in \cite{virtual_storage}. Results of the fixed price approaches are also very susceptible to the solver method. The interior point method, as opposed to the simplex method, has a tendency of finding solutions with intermediate variable values, balancing the ES (dis)charging during low- or high-price periods since there are multiple hours with the same prices. A more even charging across multiple hours results in more favourable actual profits, despite the identical computed profits for both fixed-prices methods. In contrast, the proposed bilevel model based on CPSOTA AC OPF almost perfectly matches the computed and the verified values regardless of the considered ES bus placement. The CPSOTA's accuracy, being a Taylor expansion-based model, can be iteratively even further enhanced by reevaluating the operating point parameters $\Vtn$ and $\fitn$. Regarding the system expenses, i.e. the lower-level objective function value, the Jabr's model consistently underestimates it, as the model is a relaxation. The same goes for the DC model since it does not consider network losses. The fixed-prices system expenses are computed assuming that they increase at marginal prices from a base point, i.e. systems expenses without the ES performing arbitrage. These as well underestimate the verified system expenses. Finally, the CPSOTA model results in negligible system expenses inaccuracy, indicating that the model almost perfectly computes the AC OPF.

Model accuracy is further examined on the 5\_pjm network, whose results are displayed in Table \ref{tab:net5}. Comparing the active power only bids, the Jabr's model highly overestimates the ES profit when connected to buses 1, 2 and 5. On the other hand, it underestimates the ES profit when connected to bus 4, while for the ES at bus 3 it accurately computes the ES profit. The reason for such diverse ES profit accuracy is that due to large AC OPF relaxation errors observed in the computed system expenses, wrong generators can be claimed as marginal ones, thus largely impacting the prices. This effect is very pronounced since the 5\_pjm network only has linear generator cost curves. The DC model is again relatively accurate, but overestimating the ES profit at all buses by 4-11\%. The centralized model again achieves similarly favourable ES profits as the DC model, i.e. slightly better at four busses and slightly worse at bus 5, but always achieves the lowest system expenses. Both fixed-price models also highly overestimate the ES profits, regardless on the ES position. This is because these models do not consider the effect the ES bidding strategy has on market prices. Namely, the ES tends to increase market prices when purchasing energy and reduce them when selling energy. The verified profits when the ES is located at bus 5 are actually negative for both fixed-price models, which is a result of ignoring the price changes by the ES's bidding actions. The proposed CPSOTA model results in almost perfect ES profit accuracy at all buses. The proposed model also results in almost identical computed and verified system expenses. Again, both the Jabr's and the DC models, as well as the fixed-price models, underestimate the actual system expenses.

Table \ref{tab:net5} also includes the results when the ES bids both active and reactive power. These results generally show lower accuracy than in the case when only active power is bid. Reactive power is generally more difficult to accurately model with typically a few times greater power flow inaccuracy than for active power \cite{accuracy_paper}. The Jabr's model again highly overestimates the ES profit at bus 1. At buses 2 and 3 the computed values are close to the actual ones, however, the actual profits are farther from the highest achieved with CPSOTA as compared to the active-power-only bidding. CPSOTA still achieves the highest actual ES profits. On the other hand, the centralized model is more consistent in terms of accuracy than the Jabr's model due to the use of an exact AC OPF, but has 2-4\% lower ES profits than CPSOTA due to lack of bilevel optimization structure. The actual profit increase due to reactive power bidding highly depends on the ES placement. At bus 1 the increase is 365.66 (45.4\%), at bus 2 it is 351.57 (21.3\%), at bus 3 it is 58.32 (3.0\%) and no increase at buses 4 and 5 due to zero reactive power prices.


Since the proposed CPSOTA model resulted in higher than normal inaccuracy on the 3\_lmbd network for active-power-only bids for ES placed at bus 3 (however, this inaccuracy is still extremely low, less than 0.05\%) and at buses 2 and 3 for active and reactive power bids on the 5\_pjm network (0.54\% and 0.41\% error, respectively), we ran the second iteration of the Algorithm 1 (presented in the Part I paper), whose results are presented in Table \ref{tab:net3_iter}. The second iteration basically eliminates the remaining errors, displaying fast iterative convergency of the proposed algorithm. Thus, if an extremely high accuracy is required, this can be achieved by running the second iteration, which brings the error virtually to zero since the greatest remaining ES profit error is less than 0.02\%.

\begin{table}[b]
    \setlength{\tabcolsep}{2pt} 
    \renewcommand{\arraystretch}{1} 
    \caption{CPSOTA model accuracy in the second iteration}
\label{tab:net3_iter}
\centering
\scriptsize
\begin{tabular}{lc|ccc|ccc|}
\cline{3-8}
                                              &           & \multicolumn{3}{c|}{ES profit} & \multicolumn{3}{c|}{System expenses} \\ \hline
\multicolumn{1}{|l|}{Network}                 & ES at bus & Actual       &  Computed   & \multicolumn{1}{c|}{\begin{tabular}[c]{@{}c@{}} Diff\\  {[}\%{]}\end{tabular}}   & Actual          & Computed  & \multicolumn{1}{c|}{\begin{tabular}[c]{@{}c@{}} Diff\\  {[}\%{]}\end{tabular}}        \\ \hline
\multicolumn{8}{|c|}{ES active-power-only bids}                                                                                      \\ \hline
\multicolumn{1}{|l|}{3\_lmbd}                 & 3         & 2016.876       & 2016.884  &  4.0e-4    & 98497.667  &  98497.666    &  -1.0e-6    \\ \hline
\multicolumn{8}{|c|}{ES active and reactive power bids}                                                                              \\ \hline
\multicolumn{1}{|l|}{\multirow{2}{*}{5\_pjm}} & 2         & 1999.662       & 1999.651  &  -5.5e-4    & 294728.228        & 294728.227  &  -3.4e-7     \\
\multicolumn{1}{|l|}{}                        & 3         & 2017.189       & 2016.792   &  -0.020   & 294708.966        & 294708.974   &  2.7e-6    \\ \hline
\end{tabular}
\end{table}

Statistical data for larger networks obtained by running the CPSOTA-based model for ES at each bus is presented in Table \ref{tab:allnetworks}. Buses with zero reactive power prices are excluded from the statistics for bidding both the active and reactive power. This makes for a total of 422 optimizations, 232 for active-power-only bids and 190 for active and reactive power bids. The shown errors are computed as a percentage relative difference between the actual and the computed values. As the optimization is run for the ES placement at each bus, the median, the mean and the maximum (Max) errors are listed in Table \ref{tab:allnetworks}. The median errors better represent the most common error values than the mean errors since mean are significantly influenced by the outliers. The active power bidding median, as well as the mean ES profit errors, are mainly in the range 0.01\%--1e-3\% and the system expenses median and mean errors, i.e. AC OPF errors, are in the range 1e-4\%--1e-6\%. The maximum errors, which occurred at the same bus and network, when the ES is bidding only active power are 0.19\% for the ES profit and 1.7e-3\% for the system expenses, which reduces to 4.0e-3\% and 1.5e-7\% in the second iteration of Algorithm 1. Errors when the ES bids both active and reactive powers are higher. The median and mean ES profit errors mainly range from 0.10\% to 1e-3\% and system expenses mean and median errors are in the range 1e-3--1e-5\%. However, the maximum ES profit errors are significant ($\ge1\%$) on four networks, reaching 3.97\% at bus 9 of the 30\_fsr network. We run the second iteration of Algorithm 1 for that case and the error was reduced to 0.035\%. The maximum system expense error, which also occurred at the same bus and network as the maximum ES profit error, is 0.023\%. In the second iteration it is reduced to 3.3e-4\%.


\begin{table}[tb]
    \setlength{\tabcolsep}{3pt} 
    \renewcommand{\arraystretch}{1} 
    \caption{Statistics on CPSOTA bilevel accuracy in the first iteration}
    \label{tab:allnetworks}
\centering
\begin{tabular}{l|ccc|ccc|}
\cline{2-7}
                                    & \multicolumn{3}{c|}{\begin{tabular}[c]{@{}c@{}}ES profit\\ errors (\%)\end{tabular}} & \multicolumn{3}{c|}{\begin{tabular}[c]{@{}c@{}}System expenses\\ errors (\%)\end{tabular}} \\ \hline
\multicolumn{1}{|l|}{Network}       & Median    & Mean      & Max      & Median      & Mean        & Max        \\ \hline
\multicolumn{7}{|c|}{ES active-power-only bids} \\\hline
\multicolumn{1}{|l|}{3\_lmbd}       & 5.8e-3 & 0.019   & 0.047  & 7.1e-5  & 9.3e-5 & 1.7e-4 \\ \hline
\multicolumn{1}{|l|}{5\_pjm}        & 9.5e-4 & 2.6e-3 & 0.010  & 1.3e-6  & 3.0e-6 & 6.5e-6 \\ \hline
\multicolumn{1}{|l|}{14\_ieee}      & 5.7e-4 & 8.5e-4 & 3.3e-3 & 2.4e-6  & 4.4e-6 & 1.7e-5 \\ \hline
\multicolumn{1}{|l|}{24\_ieee\_rts} & 1.9e-3 & 2.3e-3 & 8.3e-3 & 2.7e-6  & 3.2e-6 & 1.1e-5 \\ \hline
\multicolumn{1}{|l|}{30\_as}        & 2.6e-3 & 4.9e-3 & 0.033  & 8.8e-6  & 1.2e-5 & 6.4e-5 \\ \hline
\multicolumn{1}{|l|}{30\_fsr}       & 2.7e-3 & 4.6e-3 & 0.026  & 7.1e-6  & 1.2e-5 & 5.1e-5 \\ \hline
\multicolumn{1}{|l|}{30\_ieee}       & 0.014 & 0.025 & 0.19  & 1.4e-4  & 2.5e-4 & 1.7e-3 \\ \hline
\multicolumn{1}{|l|}{39\_epri}      & 1.3e-3 & 3.7e-3 & 0.024  & 4.1e-7  & 9.5e-7 & 6.5e-6 \\ \hline
\multicolumn{1}{|l|}{57\_ieee}      & 3.2e-3 & 5.9e-3 & 0.027  & 2.9e-6  & 8.7e-6 & 6.0e-5 \\ \hline
\multicolumn{7}{|c|}{ES active and reactive power bids} \\\hline
\multicolumn{1}{|l|}{5\_pjm}        & 0.41      & 0.33       & 0.53     & 1.4e-3      & 1.1e-3     & 1.8e-3     \\ \hline
\multicolumn{1}{|l|}{14\_ieee}      & 4.4e-3    & 0.10       & 0.88     & 4.2e-5      & 2.4e-4     & 1.8e-3     \\ \hline
\multicolumn{1}{|l|}{24\_ieee\_rts} & 3.7e-3    & 8.5e-3    & 0.072    & 1.0e-5      & 4.5e-5    & 3.3e-4     \\ \hline
\multicolumn{1}{|l|}{30\_as}        & 2.6e-3    & 7.2e-3    & 0.043    & 2.3e-5      & 3.2e-5     & 1.1e-4     \\ \hline
\multicolumn{1}{|l|}{30\_fsr}       & 6.1e-3    & 0.22       & 3.97     & 4.9e-5      & 1.3e-3   & 0.023      \\ \hline
\multicolumn{1}{|l|}{30\_ieee}       & 0.034    & 0.13       & 1.68     & 8.8e-4      & 1.5e-3   & 1.6e-3      \\ \hline
\multicolumn{1}{|l|}{39\_epri}      & 0.017     & 0.064      & 1.12     & 4.3e-6      & 2.9e-5     & 5.1e-4     \\ \hline
\multicolumn{1}{|l|}{57\_ieee}      & 0.087     & 0.18     & 1.28     & 1.4e-4      & 2.5e-4   & 1.3e-3      \\ \hline
\end{tabular}
\end{table}

\subsection{Case Study II: Economical Benefits of ES Reactive Power Bids}
\label{sub:economical}

\begin{figure}[b]
  \centering
  \includegraphics[scale=0.675,trim={0.65cm 0cm 1.1cm 0.3cm},clip]{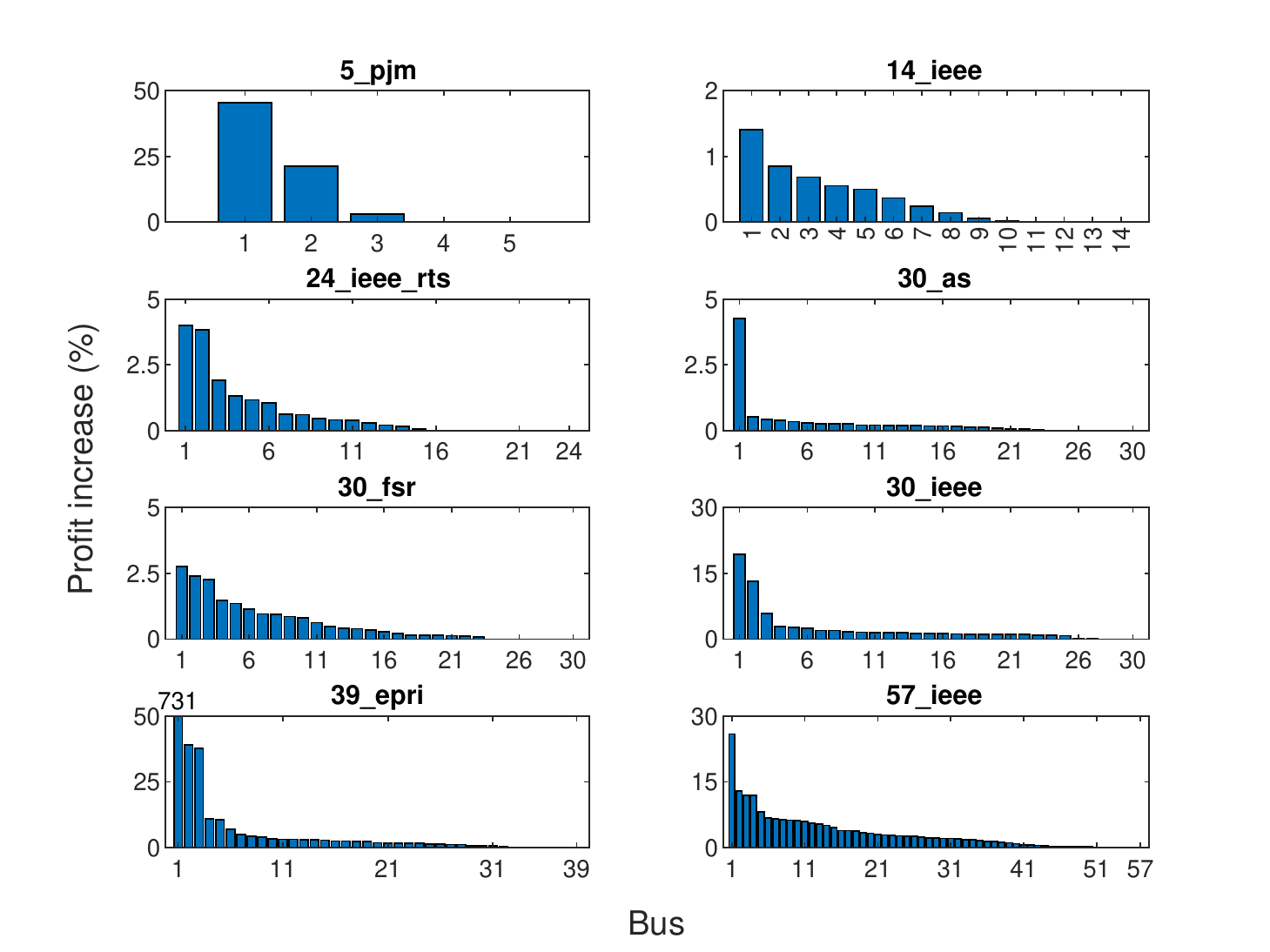}
  \caption{Relative profit increase due to reactive power bids, in descending order.}
  \label{fig:ES_profit}
\end{figure}

Reactive power bids provide both the financial opportunity for the ES and benefit for the system. Reactive power prices are commonly 10 to 100 times lower than active power prices since generators can produce them without costs, leaving only indirect active power savings to influence the price. However, since the reactive power does not consume the ES state-of-energy, but only its power capacity, it can bid it in large quantities. 3\_lmbd network is dropped from the following analysis since it has zero reactive power prices at all buses and time periods so the ES profit increase and the system savings are 0. Figure \ref{fig:ES_profit} shows the percentage profit increase for the ES due to reactive power bids sorted in a descending order. The profit increase significantly depends on the network and ES placement. At 5\_pjm, 39\_epri and 57\_ieee networks the highest ES profit increases are in the range 26\%--47\% with one outlining profit increase of 731\% at bus 30 of the 39\_epri network. This large relative increase is a result of a low active power profit due to the constant active power marginal prices at this bus caused by a large generator, i.e. the ES just discharges all the stored energy and performs no other arbitrage. On the other hand, high locational reactive power prices are a result of a large generator with high minimum reactive power output connected to bus 30. At most of the other buses and  considered networks, the profit increases are much lower and range from 0\% to 5\% as shown in the Figure \ref{fig:ES_profit}.

\begin{figure}[t]
  \centering
  \includegraphics[scale=0.675,trim={0.65cm 0cm 1.1cm 0.3cm},clip]{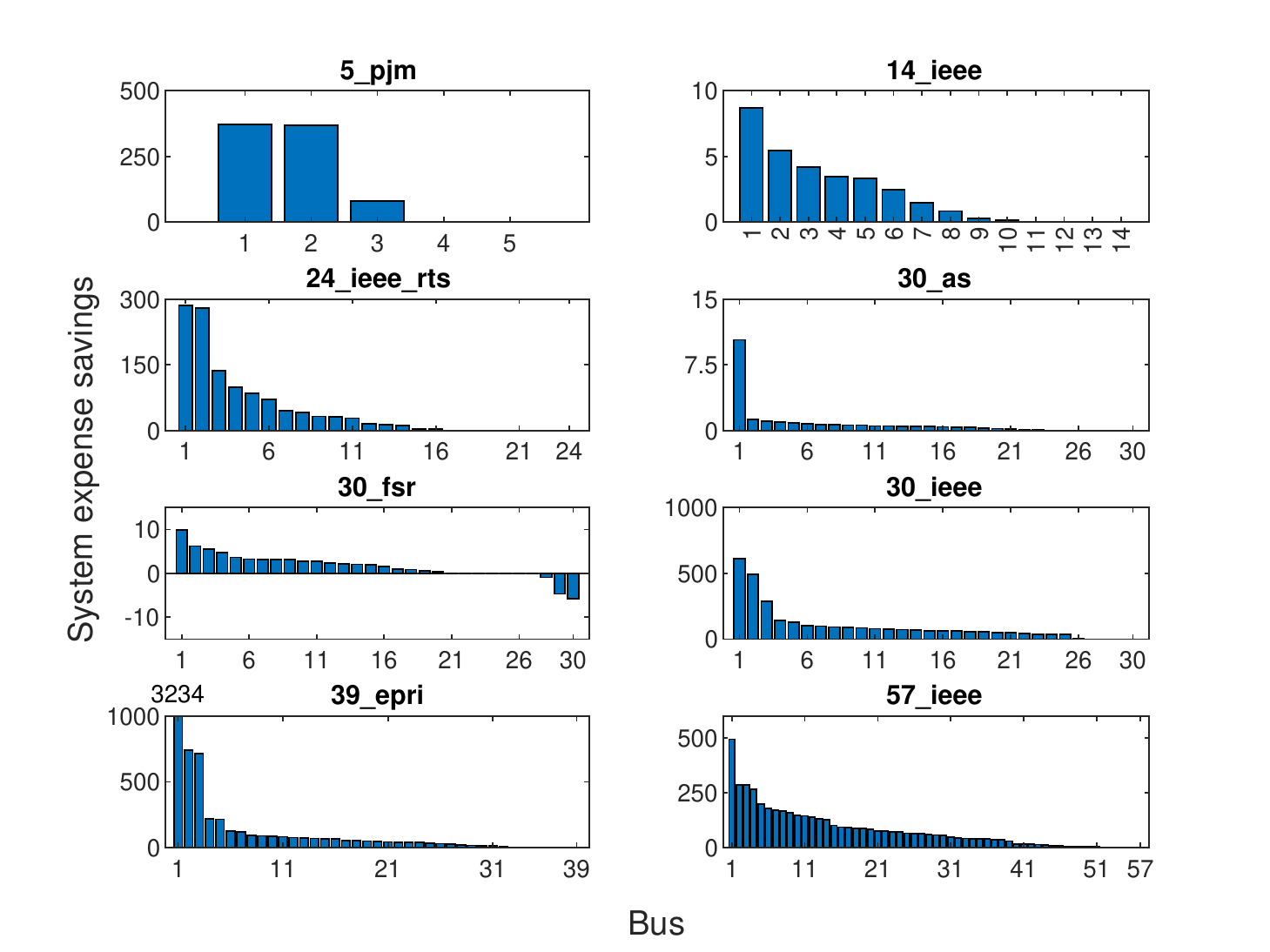}
  \caption{Absolute system expense savings due to ES reactive power bids, in descending order.}
  \label{fig:SE_savings}
\end{figure}

As the dual lower-level objective function $\Omegad$, i.e. system expenses, contains the upper-level profit term with a negative sign, $-\sum_{t,i\in \BB} (\qbat \cdot \lambdaJedan + \qqbat \cdot \lambdaDva)$, the upper-level profit increase normally results in system savings, thus the interests of the ES and the system generally align. In Figure \ref{fig:SE_savings} we see that the savings are of similar distribution as the profit increases from Figure \ref{fig:ES_profit}. However, ES savings on average result in even grater system savings, as shown in Table \ref{tab:ratio}. The ratio of absolute savings and profit increase also depends on the network, however, it mostly ranges from 1.4 to 1.6. On rare cases, the ES reactive power bids can be counter productive for the system. Figure \ref{fig:ES_profit} shows that at three buses at 30\_fsr network, the system expenses have increased. The magnitude of the increase is, however, too low to be of significance (note that Figure \ref{fig:SE_savings} contains absolute values and not percentages).

\begin{table}[t]
\caption{Average absolute system expense savings to average ES profit increase ratio due to reactive power bids}
    \label{tab:ratio}
    \setlength{\tabcolsep}{0.7pt} 
    
\begin{tabular}{|l|cccccccc|}
\hline
Network &  5\_pjm & 14\_ieee & 24\_ieee\_rts & 30\_as & 30\_fsr & 30\_ieee & 39\_epri & 57\_ieee \\ \hline
Ratio   & 1.06   & 1.59     & 1.48          & 1.53   & 1.57 & 1.44   & 1.36     & 1.50     \\ \hline
\end{tabular}
\end{table}

\subsection{Case Study III: Solution Techniques Study}
\label{sub:tractability}

\begin{table}[htb]
    \setlength{\tabcolsep}{3pt} 
    \renewcommand{\arraystretch}{1} 
    \caption{Techniques comparison on 3\_lmbd network}
    \label{tab:tab5}
    \centering
    \scriptsize
\begin{tabular}{l|cccccc|}
\cline{2-7}
                                        & \multicolumn{6}{c|}{ES at bus 3}                                                                                                                                                                                                                                       \\ \cline{2-7} 
                                        & \multicolumn{3}{c|}{ES profit}                                                                   & \multicolumn{1}{c|}{\begin{tabular}[c]{@{}c@{}}System\\ expenses\end{tabular}}      & \multicolumn{2}{c|}{Numerical tractability}                                   \\ \hline
\multicolumn{1}{|l|}{Technique}            & Actual  & Computed & \multicolumn{1}{c|}{\begin{tabular}[c]{@{}c@{}} Diff\\  {[}\%{]}\end{tabular}} & \multicolumn{1}{c|}{\begin{tabular}[c]{@{}c@{}}Duality\\ gap {[}\%{]}\end{tabular}} & Time {[}s{]} & \begin{tabular}[c]{@{}c@{}}Iterations /\\ MIP gap\end{tabular} \\ \hline
\multicolumn{1}{|l|}{PD}                & 1988.07 & 2273.58 &  14                                                                           & 0.33                                                                                & 0.08         & 33                                                             \\
\multicolumn{1}{|l|}{PD-S}              & 2013.94 & 2094.55 &  4.0                                                                          & 8.6e-2                                                                              & 0.06         & 20                                                             \\ \hline
\multicolumn{1}{|l|}{MC}                & 1988.06 & 2273.58 &  14                                                                           & 0.33                                                                                & 0.09         & 35                                                             \\ \hline
\multicolumn{1}{|l|}{SD}                & 2016.84 & 2019.07 &  0.11                                                                         & 5.2E-13                                                                             & 2.0          & 440                                                            \\
\multicolumn{1}{|l|}{SD-R $\eps$=0.1}   & \multicolumn{4}{c}{Converged to infeas. point}                                                                                                                                         & 11           & 1734                                                           \\
\multicolumn{1}{|l|}{SD-R $\eps$=1}     & 2016.75 & 2046.98 &  1.5                                                                          & 1.0E-03                                                                             & 2.3          & 396                                                            \\
\multicolumn{1}{|l|}{SD-R $\eps$=10}    & 2016.18 & 2103.68 &  4.3                                                                          & 0.010                                                                               & 0.34         & 76                                                             \\ \hline
\multicolumn{1}{|l|}{BE-SD}             & 1484.62 & 1519.93 &  2.4                                                                          & 2.1E-04                                                                             & 1800         & 51.93\%                                                        \\
\multicolumn{1}{|l|}{UE-SD}             & \multicolumn{4}{c}{No solution found}                                                                                                                                                  & 1800         & -                                                              \\ \hline
\multicolumn{1}{|l|}{BE-PF $\pen$=100}  & 1250.53 & 1291.77 &  3.3                                                                          & 2.9E-05                                                                             & 1800         & (2.8e+5)\%                                                     \\
\multicolumn{1}{|l|}{UE-PF $\pen$=100}  & -379.88 & -361.00 &  -5.0                                                                         & 4.7E-05                                                                             & 1800         & (1.2e+6)\%                                                     \\ \hline
\multicolumn{1}{|l|}{PF-SD $\pen$=10}   & 2016.69 & 2038.54 &  1.1                                                                          & 1.9E-03                                                                             & 0.17         & 40                                                             \\
\multicolumn{1}{|l|}{PF-SD $\pen$=30}   & 2016.81 & 2025.13 &  0.41                                                                         & 2.4E-04                                                                             & 0.21         & 50                                                             \\
\multicolumn{1}{|l|}{PF-SD $\pen$=100}  & 2016.82 & 2020.00 &  0.16                                                                         & 2.3E-05                                                                             & 0.10         & 35                                                             \\
\multicolumn{1}{|l|}{PF-SD $\pen$=300}  & 1958.59 & 1958.50 &  -4.6e-3                                                                      & 7.8E-06                                                                             & 1.5          & 238                                                            \\
\multicolumn{1}{|l|}{PF-SD $\pen$=1000} & 2015.15 & 2015.93 &  0.039                                                                        & 5.5E-07                                                                             & 2.4          & 334                                                            \\ \hline
\multicolumn{1}{|l|}{PF-CS $\pen$=3}    & 2016.72 & 2036.52 &  0.98                                                                         & 4.8E-03                                                                             & 0.21         & 40                                                             \\
\multicolumn{1}{|l|}{PF-CS $\pen$=10}   & 2016.81 & 2024.56 &  0.38                                                                         & 6.3E-04                                                                             & 0.16         & 42                                                             \\
\multicolumn{1}{|l|}{PF-CS $\pen$=30}   & 2016.83 & 2020.18 &  0.17                                                                         & 7.9E-05                                                                             & 0.65         & 134                                                            \\
\multicolumn{1}{|l|}{PF-CS $\pen$=100}  & 2016.84 & 2018.52 &  0.083                                                                        & 7.4E-06                                                                             & 0.36         & 77                                                             \\
\multicolumn{1}{|l|}{PF-CS $\pen$=300}  & 1915.59 & 1916.60 &  0.053                                                                        & 3.9E-06                                                                             & 0.55         & 109                                                            \\ \hline
\multicolumn{1}{|l|}{CS}                & \multicolumn{4}{c}{Converged to infeas. point}                                                                                                                                         & 20           & 1152                                                           \\
\multicolumn{1}{|l|}{CS-R $\eps$=1e-4}  & 2016.84 & 2018.97 &  0.11                                                                         & 1.4E-05                                                                             & 8.6          & 864                                                            \\
\multicolumn{1}{|l|}{CS-R $\eps$=1e-3}  & 2016.84 & 2021.72 &  0.24                                                                         & 1.2E-04                                                                             & 0.26         & 55                                                             \\
\multicolumn{1}{|l|}{CS-R $\eps$=0.01}  & 2016.83 & 2029.56 &   0.63                                                                         & 1.2E-03                                                                             & 0.12         & 28                                                             \\
\multicolumn{1}{|l|}{CS-R $\eps$=0.1}   & 2016.73 & 2049.16 &  1.6                                                                          & 9.6E-03                                                                             & 0.12         & 28                                                             \\ \hline
\multicolumn{1}{|l|}{CS-A}              & 2016.84 & 2017.86 &  0.051                                                                        & 1.6E-12                                                                             & 5.2          & 974                                                            \\
\multicolumn{1}{|l|}{CS-AR $\eps$=1e-4} & 1798.02 & 1802.96 &  0.27                                                                         & 6.2E-05                                                                             & 0.79         & 172                                                            \\
\multicolumn{1}{|l|}{CS-AR $\eps$=1e-3} & 2016.81 & 2030.67 &  0.69                                                                         & 6.2E-04                                                                             & 0.82         & 186                                                            \\
\multicolumn{1}{|l|}{CS-AR $\eps$=0.01} & 2016.69 & 2054.46 &  1.9                                                                          & 6.2E-03                                                                             & 0.56         & 128                                                            \\
\multicolumn{1}{|l|}{CS-AR $\eps$=0.1}  & 2015.52 & 2093.16 &  3.9                                                                          & 6.2E-02                                                                             & 0.07         & 17                                                             \\ \hline
\multicolumn{1}{|l|}{SM1 $\eps$=1e-4}   & 2016.84 & 2017.78 &  0.047                                                                        & 8.6E-09                                                                             & 0.11         & 22                                                             \\
\multicolumn{1}{|l|}{SM1 $\eps$=1e-3}   & 2016.84 & 2017.78 &  0.047                                                                        & 5.7E-07                                                                             & 0.43         & 69                                                             \\
\multicolumn{1}{|l|}{SM1 $\eps$=0.01}   & 2016.84 & 2017.75 &  0.045                                                                        & 3.4E-05                                                                             & 0.14         & 32                                                             \\
\multicolumn{1}{|l|}{SM1 $\eps$=0.1}    & 2016.85 & 2012.52 &  -0.21                                                                        & 5.7E-03                                                                             & 0.22         & 45                                                             \\
\multicolumn{1}{|l|}{SM1 $\eps$=1}      & 2016.84 & 1487.22 &  -26                                                                          & 0.57                                                                                & 0.09         & 22                                                             \\ \hline
\multicolumn{1}{|l|}{SM2 $\eps$=1e-4}   & 2016.84 & 2017.78 &  0.047                                                                        & 5.3E-09                                                                             & 0.12         & 21                                                             \\
\multicolumn{1}{|l|}{SM2 $\eps$=1e-3}   & 2016.84 & 2017.78 &  0.047                                                                        & 4.9E-07                                                                             & 0.55         & 74                                                             \\
\multicolumn{1}{|l|}{SM2 $\eps$=0.01}   & 2016.84 & 2017.74 &  0.045                                                                        & 4.6E-05                                                                             & 0.10         & 21                                                             \\
\multicolumn{1}{|l|}{SM2 $\eps$=0.1}    & 2016.85 & 2012.52 &  -0.21                                                                        & 5.7E-03                                                                             & 0.13         & 26                                                             \\
\multicolumn{1}{|l|}{SM2 $\eps$=1}      & 2016.84 & 1487.22 &  -26                                                                          & 0.57                                                                                & 0.12         & 27                                                             \\ \hline
\end{tabular}
\end{table}

This case study evaluates effectiveness of all solution techniques from Section \ref{sec:sol_tech} on two networks. A broad analysis allowing for both the ES active and reative power bids is performed on a small 3\_lmbd network consisting of only three buses to identify viable techniques. The results of this analysis are presented in Table \ref{tab:tab5}. Only the select techniques are applied to the 24\_ieee\_rts network for ES at bus 3. We selected the 24 bus network as it is the first larger network from the benchmark library \cite{pglib} that has quadratic generators bid curves and bus 3 since it is the first bus with non zero reactive power prices. The results of this analysis are displayed in Table \ref{tab:tab6} and show the techniques performance when the ES bids only active power and both the active and reactive powers. All the simulations, except for those referring to the interaction discretization techniques, do not have binary variables to forbid simultaneous charging and discharging. Leaving out binary variables from optimization does not impact the solutions due to nonnegative energy prices. Finally, Table \ref{tab:tab7} presents an analysis for the most reduced and the best-performing set of solution techniques to solve the problem considering also the ES (dis)charging binary variables.

As displayed in Table \ref{tab:tab5}, the primal-dual counterpart (PD) and its straightened variant (PD-S) are the easiest to compute since they belong to the convex SOCP optimization class. However, they leave large duality gaps and the straightened variant is only applicable when using the generators' quadratic bid curves. The best performing techniques (observing both the accuracy and numerical tractability) are SM1 with $\eps$=1e-4, SM2 with $\eps$=1e-4, CS-R with $\eps$=0.01, PF-SD with $\pen$=100 and PF-CS with $\pen$=10. They converge in 0.1--0.2 seconds and require only 21--42 iterations, achieving small duality gaps of 8.6e-9\%, 5.3e-9\%, 1.2e-3\%, 2.3e-5\% and 6.3e-4\%, respectively. Other defacto exact techniques are strong duality (SD), discretization techniques (BE and UE) and complementary slackness (CS and CS-A). However, a significant numerical tractability issues are observed. They manifest either as a large number of solver iterations to achieve convergence or an inability to adequately close the mixed-integer programming (MIP) gap (in the case of discretization techniques). Solvers using discretization techniques also fail to adequately set the best bound, with the strong duality version setting it to the solution of the PD technique and with the penalty factor version setting it to the effective infinity, which leaves large MIP gaps. With these techniques we used 32 discretization segments. The PF-SD technique with high penalty factor ($\pen$=300 and 1000) converges to a suboptimal solution as can occur with nonconvex formulations, while the complementary slackness technique fails to find any solution. Relaxing the techniques or reducing the penalty factor enhances the numerical tractability (for SD-R, PF-SD, PF-CS, CS-R and CS-AR), but also reduces accuracy. The McCormick technique (MC), which relaxes the bilinear terms in the strong duality, achieves the same solution as if there was no relaxed strong duality constraint which means that the relaxation is too strong to be useful. For MC we used fixed envelope bounds around the operating point of $\pm$ 1000 [1/p.u.] (i.e. 10 per MW) for active power and $\pm$ 300 [1/p.u.] (i.e. 3 per MW) for reactive power price.



\begin{table}[t]
    \setlength{\tabcolsep}{3pt} 
    \renewcommand{\arraystretch}{1} 
    \caption{Comparison of select techniques on 24\_ieee\_rts network}
    \label{tab:tab6}
    \centering
    \scriptsize
\begin{tabular}{l|cccccc|}
\cline{2-7}
                                       & \multicolumn{6}{c|}{ES at bus 3}                                                                                                                                           \\ \cline{2-7} 
                                       & \multicolumn{3}{c|}{ES profit}         & \multicolumn{1}{c|}{\begin{tabular}[c]{@{}c@{}}System\\ expenses\end{tabular}}      & \multicolumn{2}{c|}{Numerical tractability} \\ \hline
\multicolumn{1}{|l|}{Technique}           & Actual  & Computed & \multicolumn{1}{c|}{\begin{tabular}[c]{@{}c@{}} Diff\\  {[}\%{]}\end{tabular}} & \multicolumn{1}{c|}{\begin{tabular}[c]{@{}c@{}}Duality\\ gap {[}\%{]}\end{tabular}} & Time {[}s{]}          & Iterations          \\ \hline
\multicolumn{7}{|c|}{ES active-power-only bids} \\\hline
\multicolumn{1}{|l|}{PD}                & 4841.37   & 4921.32         &  1.7
             & 4.3e-3                                                                             & 0.86                  & 45                  \\
\multicolumn{1}{|l|}{PD-S}              & 4841.11   & 4902.15           &  1.3
           & 3.3e-3                                                                             & 0.83                  & 42                  \\ \hline
\multicolumn{1}{|l|}{SD}                & -10452.73 & -10444.28         &  -0.081
           & 3.5e-11                                                                             & 199                   & 2563                \\
\multicolumn{1}{|l|}{SD-R $\eps$=10}    & 4843.38   & 4901.84           &  1.2
           & 8.1e-4                                                                               & 948                   & 8102                \\ \hline
\multicolumn{1}{|l|}{PF-SD $\pen$=10}   & 4848.66   & 4860.68           &  0.25
           & 7.8e-5                                                                             & 36                    & 139                 \\
\multicolumn{1}{|l|}{PF-CS $\pen$=10}   & 4848.81   & 4854.19           &  0.11           & 8.4e-5                                                                             & 14                    & 134                 \\ \hline
\multicolumn{1}{|l|}{CS}                & \multicolumn{3}{c}{Converged to infeas. point}   &                                                                             & 475                   & 852                 \\
\multicolumn{1}{|l|}{CS-R $\eps$=0.1}   & 4848.79   & 4880.99           &  0.66           & 9.3e-4                                                                             & 4.0                   & 67                  \\ \hline
\multicolumn{1}{|l|}{CS-A}              & \multicolumn{3}{c}{Converged to infeas. point}    &                                                                            & 400                   & 3296                \\
\multicolumn{1}{|l|}{CS-AR $\eps$=0.01} & 4841.10   & 4902.14       &  1.3               & 3.3e-3                                                                             & 2.8                   & 46                  \\ \hline
\multicolumn{1}{|l|}{SM1 $\eps$=1e-4}   & 4848.98   & 4848.88         &  -2.1e-3             & 1.2e-6                                                                             & 10                    & 93                  \\
\multicolumn{1}{|l|}{SM2 $\eps$=1e-4}   & 4848.89   & 4848.97       &  1.6e-3               & 4.2e-9                                                                             & 5.5                   & 55                  \\ \hline
\multicolumn{7}{|c|}{ES active and reactive power bids} \\\hline
\multicolumn{1}{|l|}{PD}               & 4989.28 & 5271.89         &  5.7             & 2.2e-2                                                                             & 0.91                  & 49                  \\
\multicolumn{1}{|l|}{PD-S}             & 4988.72 & 5251.51         &  5.3             & 2.1e-2                                                                             & 0.83                  & 43                  \\ \hline
\multicolumn{1}{|l|}{SD}               & \multicolumn{3}{c}{Converged to infeas. point}     &                                                                         & 367                   & 4870                \\
\multicolumn{1}{|l|}{SD-R $\eps$=10}   & 5031.49 & 5118.21         &  1.7             & 8.1e-4                                                                             & 104                   & 276                 \\ \hline
\multicolumn{1}{|l|}{PF-SD $\pen$=10}  & 5034.90 & 5056.21          &  0.42            & 1.5e-4                                                                             & 62                    & 105                 \\
\multicolumn{1}{|l|}{PF-CS $\pen$=10}  & 5035.12 & 5051.37          &  0.32            & 1.2e-4                                                                             & 85                    & 138                 \\ \hline
\multicolumn{1}{|l|}{CS}               & \multicolumn{3}{c}{Converged to infeas. point} &                                                                             & 682                   & 138                 \\
\multicolumn{1}{|l|}{CS-R $\eps$=0.1}  & 5034.79 & 5070.47       &  0.71               & 5.6e-4                                                                             & 72                    & 743                 \\ \hline
\multicolumn{1}{|l|}{CS-A}             & 5035.14 & 5053.21        &  0.36              & 3.9e-10                                                                             & 599                   & 5329                \\
\multicolumn{1}{|l|}{CS-AR $\eps$=0.1} & 4988.77 & 5251.44        &  5.3              & 2.1e-2                                                                             & 40                    & 180                 \\ \hline
\multicolumn{1}{|l|}{SM1 $\eps$=1e-4}  & 5035.02 & 5035.54        &  0.010              & 5.1e-9                                                                             & 10                    & 95                  \\
\multicolumn{1}{|l|}{SM2 $\eps$=1e-4}  & 5035.02 & 5035.54        &  0.010              & 4.4e-9                                                                             & 13                    & 112                 \\ \hline
\end{tabular}
\end{table}

The select versions of the primal-dual, strong duality, penalty factor, complementary slackness and the two smoothing techniques are tested on a larger network with results displayed in Table \ref{tab:tab6}. The previously well performing techniques PF-SD and PF-CS (see Table \ref{tab:tab5}) did not perform well. In the case of bidding both the active and reactive powers it took 62 and 85 second to compute them, which is about 6 times longer than for the two smoothing techniques and about 3 times longer when only active power bidding is allowed. CS-R also takes 6 times longer to compute than the smoothing techniques in case of bidding both the active and reactive powers, but it finishes slightly faster then the smoothing techniques when only active power bidding is allowed. It closes the duality gap only moderately well with 5.6e-4\% in first case and 9.3e-4\% in the latter. CS-AR performs similar to CS-R, with CS-AR being somewhat faster, but also with larger duality gaps. The primal-dual counterpart techniques still offer the best tractability, but also low accuracy due to high duality gaps (ES profit errors are approximately 5\% for active and reactive power bids and 2\% for active-power-only bids). The two smoothing techniques perform reasonably tractable in both cases finishing in 10 seconds and taking 89 iterations on average. Meanwhile, they achieve close to zero (order of magnitude 1e-8\%) duality gaps. The remaining techniques SD, CS and CS-A all either converge to an infeasible point or display serious numerical tractability issues.


\begin{table}[t]
    \setlength{\tabcolsep}{2pt} 
    \renewcommand{\arraystretch}{1} 
    \caption{Performance in discrete optimization}
    \scriptsize
    \label{tab:tab7}
    \centering
\begin{tabular}{ll|cccccc}
\cline{3-8}
                                                                                                       &                 & \multicolumn{3}{c|}{ES profit}         & \multicolumn{1}{c|}{\begin{tabular}[c]{@{}c@{}}System\\ expenses\end{tabular}}      & \multicolumn{2}{c|}{Numerical tractability} \\ \hline
\multicolumn{1}{|l|}{Network}                                                                          & Technique          & Actual  & Computed & \multicolumn{1}{c|}{\begin{tabular}[c]{@{}c@{}} Diff\\  {[}\%{]}\end{tabular}} & \multicolumn{1}{c|}{\begin{tabular}[c]{@{}c@{}}Duality\\ gap {[}\%{]}\end{tabular}} & Time {[}s{]}                       & \multicolumn{1}{c|}{Nodes}                       \\ \hline
\multicolumn{8}{|c|}{ES active-power-only bids}                                                                                                                                                                                                                                                                                                 \\ \hline
\multicolumn{1}{|l|}{\multirow{2}{*}{\begin{tabular}[c]{@{}l@{}}24\_ieee\_rts\\ (bus 3)\end{tabular}}} & SM1 $\eps$=1e-4 & 4848.98 & 4850.73       &  0.036               & 1.4e-4                                                                             & 31                                 & \multicolumn{1}{c|}{1}                           \\
\multicolumn{1}{|l|}{}                                                                                 & SM2 $\eps$=1e-4 & 4848.85 & 4848.97            &  2.5e-3          & 6.9e-9                                                                             & 15                                 & \multicolumn{1}{c|}{1}                           \\ \hline
\multicolumn{8}{|c|}{ES active and reactive power bids}                                                                                                                                                                                                                                                                                         \\ \hline
\multicolumn{1}{|l|}{\multirow{2}{*}{\begin{tabular}[c]{@{}l@{}}3\_lmbd\\ (bus 3)\end{tabular}}}       & SM1 $\eps$=1e-4 & 2016.89 & 2017.78      &  0.044                & 5.7e-9                                                                             & 1.3                                & \multicolumn{1}{c|}{3}                           \\
\multicolumn{1}{|l|}{}                                                                                 & SM2 $\eps$=1e-4 & 2016.89 & 2017.78       &  0.044               & 5.7e-9                                                                             & 0.19                               & \multicolumn{1}{c|}{1}                           \\ \hline
\multicolumn{1}{|l|}{\multirow{2}{*}{\begin{tabular}[c]{@{}l@{}}24\_ieee\_rts\\ (bus 3)\end{tabular}}} & SM1 $\eps$=1e-4 & 5035.02 & 5035.54        &  0.010              & 4.7e-9                                                                             & 93                                 & \multicolumn{1}{c|}{3}                           \\
\multicolumn{1}{|l|}{}                                                                                 & SM2 $\eps$=1e-4 & 5035.05 & 5036.23                &  0.023      & 5.5e-5                                                                             & 71                                 & \multicolumn{1}{c|}{3}                           \\ \hline
\end{tabular}
\end{table}

This case study is completed by evaluating the performance of the two smoothing techniques with included binary variables that forbid simultaneous ES charging and discharging, as displayed in Table \ref{tab:tab7}. Both smoothing techniques achieve comparable and high accuracy. The displayed tractability of SM2 technique is marginally better.

\section{Conclusion}\label{sec:conclusion} 

The approach presented in Part I and Part II papers avoids the lower level linearization and can be used to effectively solve a strategic energy storage bilevel transmission-network-constrained market participation problem. Both active and reactive power bids are considered. The model utilises a convex polar second-order Taylor approximation \cite{cpsota} of AC OPF in the lower level thanks to which the KKT-based single-level reduction is possible while achieving extremely high AC OPF accuracy. The resulting complementary conditions are transformed using the smoothing technique to achieve numerical tractability.

Results indicate very high and consistent model accuracy tested on eight meshed transmission system networks for ES placement at every bus. For active-power-only bids, the mean and median upper-level profit errors are mainly in the range 0.01\%--1e-3\% with the maximum observed error of 0.19\% within 232 optimizations. However, this error is reduced to 4.0e-3\% in the second iteration of the Algorithm 1. When both the active and reactive power bids are considered, the upper-level profit errors are slightly higher, but still very low and mainly in the range 0.10\%--1e-3\% with the maximum observed error 3.97\% within 190 optimizations. Again, the high errors can be further reduced by iteratively running the algorithm. Already the second iteration reduces the 3.97\% error down to 0.035\%. The lower-level objective function mean and median errors (AC OPF errors) are mainly in range the 1e-3\%--1e-6\%.

Economical benefits of ES reactive power bids significantly depend on the network and ES bus placement. The highest ES profit increases are in between 26\%--47\%, while for the majority of cases up to 5\%. ES profit increase normally also reduces the system expenses, but at a greater amount. The average ES profit increase and average system savings ratio mostly ranges from 1.4--1.6.

The smoothing techniques achieve close-to-zero duality gaps, i.e. in the range 1e-6\%--1e-8\%, while outperforming in terms of tractability all other classical KKT-based single-level duality gap closure-enforcing reduction techniques.

The presented approach is also applicable to the various upper-level problems, e.g. generator, load or aggregator bidding or investment problems. It is also applicable for bilevel reserve procurement problems considering reactive power. Finally, it should benefit the system operators to assess the effect of their network investments, e.g. lines or energy storage, on the social welfare. Thus, utilization of this tool may benefit the market operators to achieve a revenue adequate and a more complete and fair market design.  

\bibliographystyle{IEEEtran}

\end{document}